\documentclass[11pt,a4paper]{article}

\usepackage[T1]{fontenc}
\usepackage[utf8]{inputenc}
\usepackage[margin=2.5cm]{geometry}
\usepackage{amsmath,amssymb}
\usepackage{graphicx}
\usepackage{array,booktabs,longtable}
\usepackage{placeins}
\usepackage[numbers,sort&compress]{natbib}
\usepackage[hidelinks]{hyperref}
\usepackage{caption}

\title{Leaky protection reveals three patterns in the relative advantages of vaccination strategies across transmission intensity}

\author{
Yiwen Xing\textsuperscript{a},
Jianyi Zhang\textsuperscript{a},
Haoliang Cui\textsuperscript{a},
Zijie Yang\textsuperscript{b},\\
Qianning Wang\textsuperscript{a},
Lang Qin\textsuperscript{a},
Zheqi Wang\textsuperscript{a,1},
Zhongwei Jia\textsuperscript{a,1}\\[0.5em]
{\small\parbox{0.9\textwidth}{\centering
\textsuperscript{a}Department of Global Health,
School of Public Health, Peking University,
Beijing, 100191, China\\[0.3em]
\textsuperscript{b}Shenzhen Center for Disease Control and Prevention, Guangdong Shenzhen, 518055, China
}}
}

\date{}

\begin{document}
\maketitle

\begin{abstract}
Targeted vaccination often prioritizes individuals who contribute most to transmission. However, this rationale may change under leaky protection, whereby vaccination reduces susceptibility at each exposure rather than fully preventing infection. Here, we use a survey-derived individual-level exposure framework for 2,279 residents to simulate infection dynamics across ranges of transmission intensity and vaccine efficacy. We compare two vaccination strategies: the high-exposure strategy (prioritizing highly exposed individuals) and the oldest-first strategy (prioritizing older individuals). As transmission intensity changes, we identify three patterns in the relative advantages of these two strategies. At low transmission, the high-exposure strategy outperforms the oldest-first strategy in reducing both epidemic peak size and final infection proportion. At intermediate transmission, it lowers epidemic peak size but increases final infection proportion, revealing a trade-off between these metrics. At high transmission, differences between the strategies narrow, and the oldest-first strategy has a slight advantage for both metrics. Overall, our findings highlight the importance of considering vaccine failure mode when evaluating vaccination policies under different transmission dynamics.
\end{abstract}


\medskip
\noindent \textsuperscript{1}Corresponding authors. E-mail: xyw980217@gmail.com, wangzhq@bjmu.edu.cn

\section*{Introduction}
\noindent Two commonly used idealized models of imperfect vaccine protection against infection assume either all-or-nothing or leaky protection. Under all-or-nothing protection, a fraction of vaccinees is fully protected while the remainder receive no protection. Under leaky protection, susceptibility is reduced for all vaccinees~\cite{smith1984assessment,halloran1992interpretation,halloran2025thirty}. Classic studies of network immunization show that, when immunized individuals are assumed to be fully protected, prioritizing individuals with high degree or central network positions can improve intervention efficiency~\cite{pastor2002immunization,cohen2003efficient,yang2019efficient}. Under leaky protection, however, infection risk can accumulate across repeated exposures to infectious individuals, leaving highly exposed vaccinees at substantial risk of infection~\cite{edlefsen2014leaky}. Assumptions about vaccine protection can influence the choice of recipients. Lee et al. show that, under certain conditions, different assumptions about vaccine failure mode can change the recommended priority group, emphasizing this factor as an important consideration in the design and evaluation of vaccination strategies~\cite{lee2025vaccine}.

Another factor that influences which vaccination strategy performs best is the epidemiological objective. When vaccine supply is limited, prioritization can favour individuals with greater exposure or those at greater risk of death following infection. The former aims to reduce infections and subsequent transmission, whereas the latter focuses on protecting individuals who are more likely to die if infected. These objectives do not necessarily favour the same vaccination choices. Previous studies show that strategies that minimize infections are not necessarily those that minimize deaths~\cite{bubar2021model,matrajt2021optimizing}. Vaccination strategies therefore need to be evaluated across epidemiological metrics, rather than judged by a single metric as an indication of their overall advantage. 

In addition to these factors, transmission intensity can also reshape the relative advantages of vaccination strategies. Modelling studies show that the effect of vaccination on the population burden of infection varies with transmission intensity even when vaccine efficacy is held constant~\cite{ragonnet2015vaccination}. This variation can affect not only the extent of the benefit but also the choice of recipients. Gavish and Katriel show that, under leaky protection, prioritizing groups with lower exposure can be nearly optimal for minimizing final epidemic size at sufficiently high transmission intensity~\cite{gavish2022optimal}. These findings indicate that prioritizing highly exposed individuals is not necessarily the most effective approach to reducing the final number of infected individuals under all transmission conditions. Transmission intensity is therefore an important consideration when assessing vaccination strategies.

\begin{figure}[!t]
    \centering
    \includegraphics[width=0.8\textwidth]{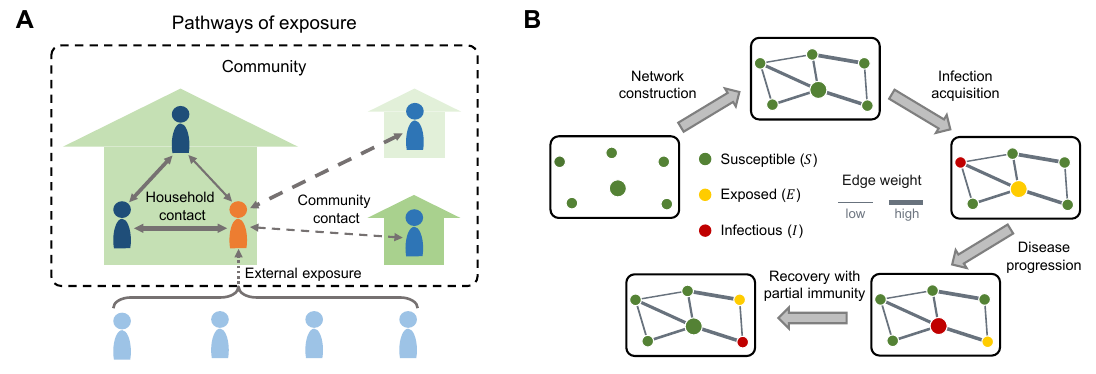}
    \caption{Schematic of the individual-level heterogeneous exposure framework. ({\textit A}) Nodes representing individuals experience exposure through two broad pathways, internal network exposure and external exposure. Internal network exposure comprises stable, higher-intensity household contacts among co-residents (solid bidirectional arrow) and stochastic community contacts generated by daily activities (dashed bidirectional arrow). Both contact types are represented as weighted network edges. Work, school and non-fixed out-of-community activities are considered in the transmission process as individual-specific external exposure (dotted arrow) rather than network edges. ({\textit B}) Nodes are connected by weighted contact edges to form the internal contact network, and line shading indicates contact intensity. Colours of nodes indicate infection states. Green, yellow and red denote susceptible, exposed and infectious states, respectively. Susceptible nodes may enter the exposed state through transmission from infectious neighbours or external infection pressure and subsequently progress to the infectious state. After recovery, infectious nodes return to the susceptible state but retain partial natural immunity that varies with time since recovery.}
    \label{fig:model}
\end{figure}

Models based on group-level average contact patterns can capture differences between population groups, but do not explicitly retain individual connections or variation in contacts within groups. These individual differences warrant particular attention under leaky protection, because infection risk can still accumulate across repeated exposures to infectious individuals after vaccination. In an observational study of an incarcerated population, Lind et al. find that closer contact with infectious individuals is associated with greater infection risk and weaker estimated vaccine protection~\cite{lind2023evidence}. These findings provide empirical support for leaky protection and motivate explicitly representing differences in individual exposure conditions. Contact networks retain connections between individuals and differences in contact intensity. They capture how infection pressure depends on infectious neighbours and how protecting one individual can affect the infection risk of others~\cite{keeling2005networks}. This framework is therefore suited to examining vaccination strategies when leaky protection and individual differences in exposure jointly shape infection outcomes. Previous research under leaky protection shows that final infection proportion and epidemic peak size can respond differently to mixing between vaccinated and unvaccinated individuals~\cite{burgio2022homophily}. This study focuses on the clustering of vaccination status in the network while constraining additional associations between vaccination status and node strength during vaccine distribution. However, previous studies do not directly establish the relationship between the two infection control objectives when vaccination strategies account for individual differences in exposure. Under leaky protection, do the objectives of reducing epidemic peak size and reducing final infection proportion favour the same vaccination strategies? And how does this relationship vary with transmission intensity?

To address these questions, we construct a survey-derived individual-level exposure framework for 2,279 residents to examine vaccination strategies under leaky protection. The framework combines an internal weighted contact network comprising household and stochastic community contacts with external exposure. We simulate individual-level epidemic dynamics across broad ranges of transmission intensity and vaccine efficacy, using a COVID-like parameterization as the primary scenario. At a fixed vaccination coverage of $50\%$, we compare the random, high-exposure and oldest-first strategies. The high-exposure strategy prioritizes individuals with high exposure strength, whereas the oldest-first strategy prioritizes individuals in descending order of age. The former provides its most consistent benefit in reducing epidemic peak size, whereas the latter consistently reduces expected deaths. We also examine hybrid allocation between these two targeted strategies to assess joint changes in epidemic peak size and final infection proportion. Comparing the two strategies reveals three patterns across transmission intensities. At low transmission intensity, both infection metrics favour the high-exposure strategy. At intermediate transmission intensity, this strategy achieves a lower epidemic peak size but a higher final infection proportion than the oldest-first strategy, revealing a trade-off between the two metrics. At high transmission intensity, the differences narrow and the oldest-first strategy has a slight advantage in both metrics.

\begin{figure}[!htbp]
    \centering
    \includegraphics[width=0.8\textwidth]{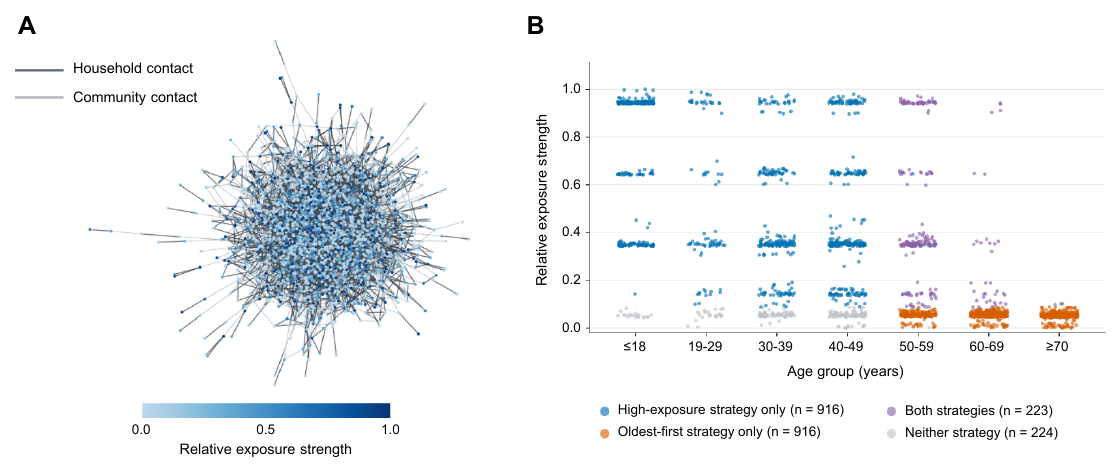}
    \caption{Individual contact network and recipients selected by the two targeted vaccination strategies. (\textit{A}) Largest connected component of the internal weighted contact network, containing 2,258 nodes (99.1\% of residents) and 4,617 edges. Each node represents a resident, with darker blue indicating higher relative exposure strength. Dark and light grey edges represent household and community contacts, respectively. Node positions do not represent geographical locations. (\textit{B}) Joint distribution of age group and relative exposure strength for all 2,279 residents. Colours distinguish recipients selected by the high-exposure strategy, the oldest-first strategy, both or neither at $50\%$ vaccination coverage. Points are randomly offset horizontally within each age group to reduce overlap. In both panels, relative exposure strength is normalized by the maximum exposure strength across all residents.}
    \label{fig:network}
\end{figure}

\section*{Results}
\subsection*{Exposure strength and age define different vaccination targets}

We construct an individual-level exposure framework for 2,279 residents using data from a community household survey. The survey records age group, household membership, work/school activity status and daily activities. Each resident is represented by a node. Household and activity information is used to define contact relationships and exposure levels. Age is used to prioritize older individuals for vaccination and to assign age-specific infection fatality ratios (IFRs), which represent mortality risk after infection. Details of survey data processing and model construction are provided in \textit{Materials and Methods} and the \textit{SI Appendix}, sections~1A--1B and Table~S1.

The framework combines an internal weighted contact network with individual external exposure (Fig.~\ref{fig:model}{\textit A}). The internal network includes contacts among household members and community contacts generated randomly using daily activity information. Edge weights represent the daily contact durations assigned in the model. Infection opportunities arising from work, school and other activities outside the network are represented by individual external infection pressure rather than network edges. This pressure is calculated from individual external exposure, external infectious prevalence and the transmission coefficient, depicting the external contribution to infection hazard (calculation provided in the \textit{SI Appendix}, section~1C). Residents can acquire infection from infectious neighbours or from outside the community network. This framework represents both individual differences in contact levels and transmission pathways through specific connections. We simulate SEIS-type transitions between susceptible (S), exposed (E) and infectious (I) states. After recovery, individuals become susceptible to infection again but retain partial immunity that varies with time since recovery (Fig.~\ref{fig:model}{\textit B}). Vaccination provides leaky protection, reducing infection probability at a given level of infection pressure without fully preventing infection.

The resulting internal weighted contact network contains 2,279 nodes and 4,624 edges, all of which are included in subsequent transmission simulations. Fig.~\ref{fig:network}{\textit A} shows its largest connected component, comprising 2,258 nodes and 4,617 edges and accounting for 99.1\% of all residents. To describe individual exposure levels, we define a node’s exposure strength as the total weight of its internal network connections plus its external exposure. This index represents relative exposure opportunities in the model rather than actual infection probability. Relative exposure strength shown in the figures is calculated by dividing this index by its maximum across all residents. This normalization does not change vaccination priority rankings.

In addition to exposure levels, we consider individual mortality risk after infection. Under the disease parameterization used here, this risk is represented by age-specific IFRs, making age another basis for vaccination priority. The joint distribution of age and exposure strength shows variation in exposure levels within each age group. High exposure strength is also not confined to a single age group (Fig.~\ref{fig:network}{\textit B}).

We therefore design and compare two targeted vaccination strategies with different priorities. The high-exposure strategy selects recipients in descending order of exposure strength, whereas the oldest-first strategy prioritizes older individuals (see \textit{Materials and Methods} for implementation details). At a vaccination coverage of 50\%, each strategy selects 1,139 residents, with 223 selected by both strategies. The high-exposure strategy mainly covers individuals with higher exposure strength across different age groups. The oldest-first strategy concentrates vaccination among older residents, many of whom have relatively low exposure strength (Fig.~\ref{fig:network}{\textit B}). The two strategies thus distribute the same number of vaccines to different but partially overlapping groups. This allows us to compare the effects of prioritizing individuals with high exposure or high mortality risk.

\begin{figure}[!t]
    \centering
    \includegraphics[width=0.8\textwidth]{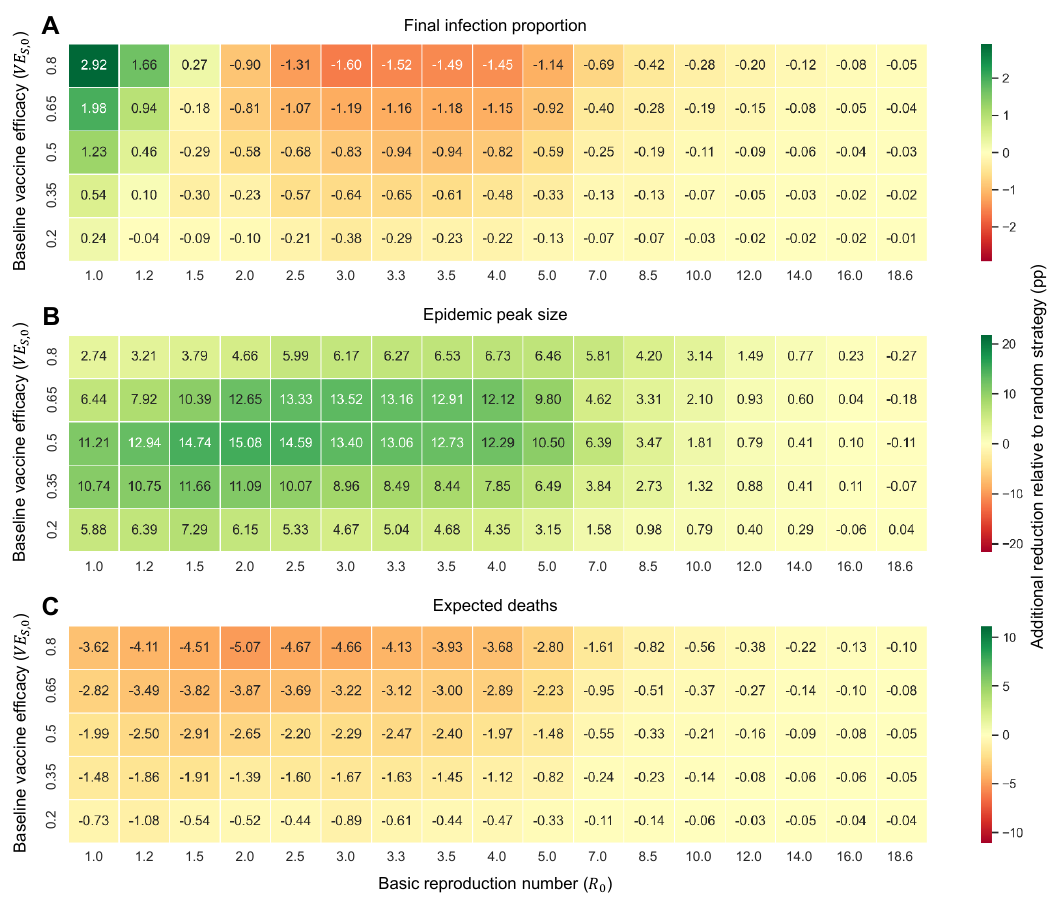}
    \caption{Additional reductions achieved by the high-exposure strategy relative to the random strategy at 50\% vaccination coverage. Heatmaps show the additional reduction across combinations of the basic reproduction number \(R_0\) and the baseline vaccine efficacy \(VE_{S,0}\) for final infection proportion (\textit{A}), epidemic peak size (\textit{B}), and expected deaths (\textit{C}). Additional reductions are expressed in percentage points using the corresponding no-vaccination mean as the normalization baseline. Positive values indicate greater reductions under the high-exposure strategy (green), whereas negative values indicate greater reductions under the random strategy (red). Compared with the random strategy, the high-exposure strategy most consistently reduces epidemic peak size, while its effects on final infection proportion vary with transmission intensity and it provides no additional reduction in expected deaths. Results are averaged over 100 Monte Carlo simulations per strategy and parameter combination. Each simulation lasts 500 days and begins with 10 randomly selected infectious individuals.}
    \label{fig:high_exposure}
\end{figure}

\subsection*{The high-exposure strategy primarily reduces epidemic peak size}

We first assess the performance of the high-exposure strategy under leaky protection. We use the random strategy as the reference, selecting vaccine recipients uniformly at random from the population. At a fixed vaccination coverage of $50\%$, we evaluate strategy performance using three epidemiological metrics: final infection proportion, epidemic peak size and expected deaths (definitions are provided in \textit{Materials and Methods}). For each metric, we calculate the additional reduction achieved by the high-exposure strategy relative to the random strategy (Fig.~\ref{fig:high_exposure}). This procedure is conducted across different combinations of two epidemic parameters: the baseline vaccine efficacy \(VE_{S,0}\), which is the initial vaccine efficacy against susceptibility, and the basic reproduction number \(R_0\), which determines the transmission intensity. Positive values indicate that the high-exposure strategy performs better than the random strategy by achieving a larger reduction, whereas negative values indicate the opposite.

The high-exposure strategy shows its most consistent benefit in reducing epidemic peak size (Fig.~\ref{fig:high_exposure}{\textit B}). Across most combinations of  $R_0$ and $VE_{S,0}$, particularly at low ($R_0\le1.2$) to intermediate ($1.5\le R_0\le12.0$) transmission intensity, this strategy achieves a greater reduction in epidemic peak size than the random strategy. The advantage is greatest when $R_0=2.0$ and $VE_{S,0}=0.5$, where the additional peak reduction is $15.08$ percentage points (pp). At the highest transmission intensities ($R_0\ge14.0$), however, the difference approaches zero and becomes slightly negative in several settings. The extreme value is $-0.27$ pp at $R_0=18.6$ and $VE_{S,0}=0.8$. These results indicate that the high-exposure strategy reduces epidemic peak size across most transmission settings, but this advantage over the random strategy diminishes and reverses in some of the highest-transmission settings.

Such a peak-suppression advantage does not consistently translate into a lower final infection proportion (Fig.~\ref{fig:high_exposure}{\textit A}). Additional reductions in final infection proportion are positive mainly at the lowest values of $R_0$, with larger benefits at high $VE_{S,0}$. When $R_0=1.0$ and $VE_{S,0}=0.8$, the high-exposure strategy achieves an additional reduction of $2.92$ pp relative to the random strategy. At low $R_0$, infections are concentrated around high-exposure nodes and their neighbouring transmission pathways, so protecting these nodes can efficiently interrupt early spread. As $R_0$ increases to intermediate values, this advantage rapidly weakens and becomes negative. The minimum is $-1.60$ pp at $R_0=3.0$ and $VE_{S,0}=0.8$. In this case, greater transmission intensity raises the infection risk associated with each infectious contact. Repeated infectious contacts can therefore leave vaccinated high-exposure individuals with a substantial cumulative probability of infection under leaky protection. This may reduce the benefit of concentrating vaccines on these individuals. By contrast, the random strategy also vaccinates individuals with lower exposure strength, for whom leaky protection may still effectively prevent infection. Thus, at intermediate transmission intensity, the high-exposure strategy may outperform the random strategy in reducing epidemic peak size but not final infection proportion. As $R_0$ increases further to high values, rising overall infection pressure weakens the influence of vaccination prioritization on final infection proportion, causing the difference between the two strategies to gradually narrow towards zero. This non-monotonic pattern may reflect the accumulation of infection risk across repeated exposures under leaky protection.

The high-exposure strategy provides no additional reduction in expected deaths relative to the random strategy (Fig.~\ref{fig:high_exposure}{\textit C}). It is outperformed by the random strategy in all scenarios, with the additional reduction in expected deaths reaching a minimum of $-5.07$ pp at $R_0=2.0$ and $VE_{S,0}=0.8$. Additionally, the contrast between improved peak control and poorer performance for expected deaths also appears under the alternative disease parameterizations (\textit{SI Appendix}, Fig. S6). In our network, many older individuals with higher age-specific IFRs have low exposure strength, so vaccination based solely on strength provides insufficient coverage for them. Consequently, when the objective shifts from reducing epidemic peak size to reducing mortality burden, the high-exposure strategy no longer retains an advantage.

\begin{figure}[!t]
    \centering
    \includegraphics[width=0.8\textwidth]{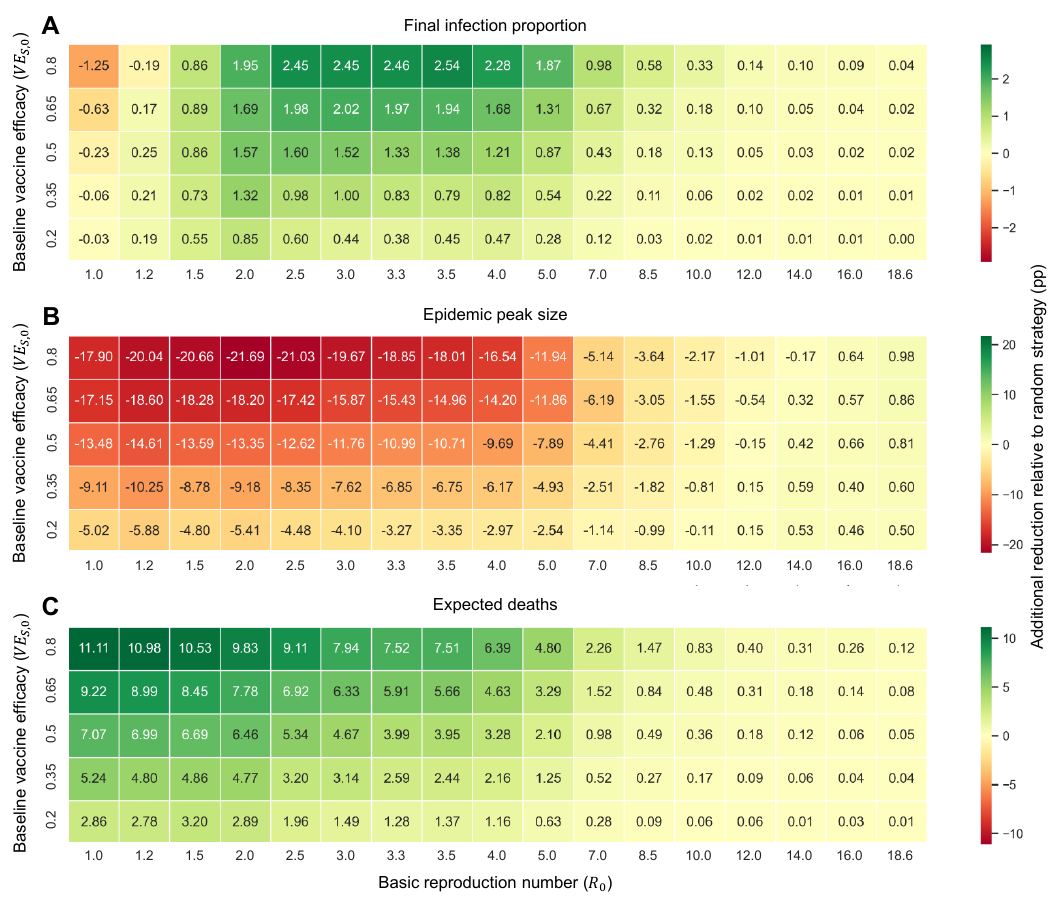}
    \caption{Additional reductions achieved by the oldest-first strategy relative to the random strategy at 50\% vaccination coverage. Heatmaps show the additional reduction across combinations of \(R_0\) and  \(VE_{S,0}\) for final infection proportion (\textit{A}), epidemic peak size (\textit{B}) and expected deaths (\textit{C}). Additional reductions are expressed in percentage points using the corresponding no-vaccination mean as the normalization baseline. Positive and negative values indicate greater (green) and smaller (red) reductions under the oldest-first strategy, respectively. Compared with the random strategy, the oldest-first strategy consistently reduces expected deaths and lowers final infection proportion in most scenarios, but generally produces a larger epidemic peak size. Results are averaged over 100 Monte Carlo simulations per strategy and parameter combination. Each simulation lasts 500 days and begins with 10 randomly selected infectious individuals.}
    \label{fig:oldest_first}
\end{figure}

\subsection*{The oldest-first strategy primarily reduces final infection proportion and expected deaths}

Unlike the high-exposure strategy, the oldest-first strategy represents a vulnerability-oriented logic. It prioritizes older individuals with higher age-specific IFRs. We again calculate the additional reduction achieved by this strategy relative to the random strategy at $50\%$ vaccination coverage (Fig.~\ref{fig:oldest_first}), comparing its performance across the three epidemiological metrics.

The oldest-first strategy performs best at reducing expected deaths (Fig.~\ref{fig:oldest_first}{\textit C}). Across all $R_0$ and $VE_{S,0}$ combinations, this strategy produces a greater reduction in mortality burden than the random strategy. The advantage is strongest under higher $VE_{S,0}$ and low transmission intensity, reaching $11.11$ pp when $VE_{S,0}=0.8$ and $R_0=1.0$. Although this advantage weakens as $R_0$ increases, it remains positive throughout the parameter space (a similar mortality advantage also appears under the alternative disease parameterizations in \textit{SI Appendix}, Fig.~S7). This indicates that, under leaky protection, directly protecting individuals with higher age-specific IFRs can robustly reduce expected deaths.

This mortality benefit is accompanied by a larger epidemic peak size in most scenarios. For epidemic peak size (Fig.~\ref{fig:oldest_first}{\textit B}), the oldest-first strategy performs worse than the random strategy in most parameter combinations, with the largest disadvantages occurring under high $VE_{S,0}$ and low-to-intermediate $R_0$. The most negative additional peak reduction is $-21.69$ pp at $VE_{S,0}=0.8$ and $R_0=2.0$. At high $R_0$ values, however, the oldest-first strategy edges out the random strategy. Here, leaky protection may be less effective at preventing infection among highly exposed individuals. In our network, many of the older individuals do not regularly attend work or school and therefore have lower exposure strength. Concentrating vaccination among them may therefore reduce their contribution to the epidemic peak more effectively than random vaccination. These results show that, although prioritizing older individuals reduces expected deaths, the oldest-first strategy generally provides weaker control of epidemic peak size than the random strategy, possibly because unvaccinated high-exposure individuals can become infected early and transmit infection to their neighbours.

The additional reduction in final infection proportion (Fig.~\ref{fig:oldest_first}{\textit A}) also shows a non-monotonic pattern as $R_0$ increases. At the lowest transmission intensity, this strategy performs worse than the random strategy. At $VE_{S,0}=0.8$ and $R_0=1.0$, its additional reduction is $-1.25$ pp. In low $R_0$ scenarios, infection spread remains more localized and is strongly shaped by high-exposure nodes. The disadvantage of the oldest-first strategy may arise because its recipients have lower exposure levels overall. As $R_0$ increases, the relative effect of the oldest-first strategy on final infection proportion becomes positive and reaches the largest advantage when $VE_{S,0}=0.8$ and $R_0=3.5$. The corresponding additional reduction is $2.54$ pp relative to the random strategy. One possible explanation is that individuals with low or intermediate exposure experience lower cumulative infection pressure and may be more likely to remain uninfected after vaccination. Therefore, when $R_0$ becomes high enough to threaten a substantial portion of the population but not so high that nearly all nodes are likely to become infected, protecting lower- or intermediate-exposure vulnerable individuals may generate a benefit in final infection proportion. For high $R_0$ values, the effect of vaccination prioritization on final infection proportion weakens as the overall infection probability increases, and the advantage of the oldest-first strategy relative to the random strategy contracts.

Across the tested network structures, the high-exposure strategy retains an advantage in epidemic peak size and the oldest-first strategy retains an advantage in expected deaths relative to the random strategy. However, the high-exposure strategy also achieves positive mean additional reductions in final infection proportion and expected deaths in the BA network (\textit{SI Appendix}, Fig. S8).

\clearpage

\begin{figure}[p]
    \centering

    \includegraphics[
        width=0.95\linewidth,
        height=0.50\textheight,
        keepaspectratio
    ]{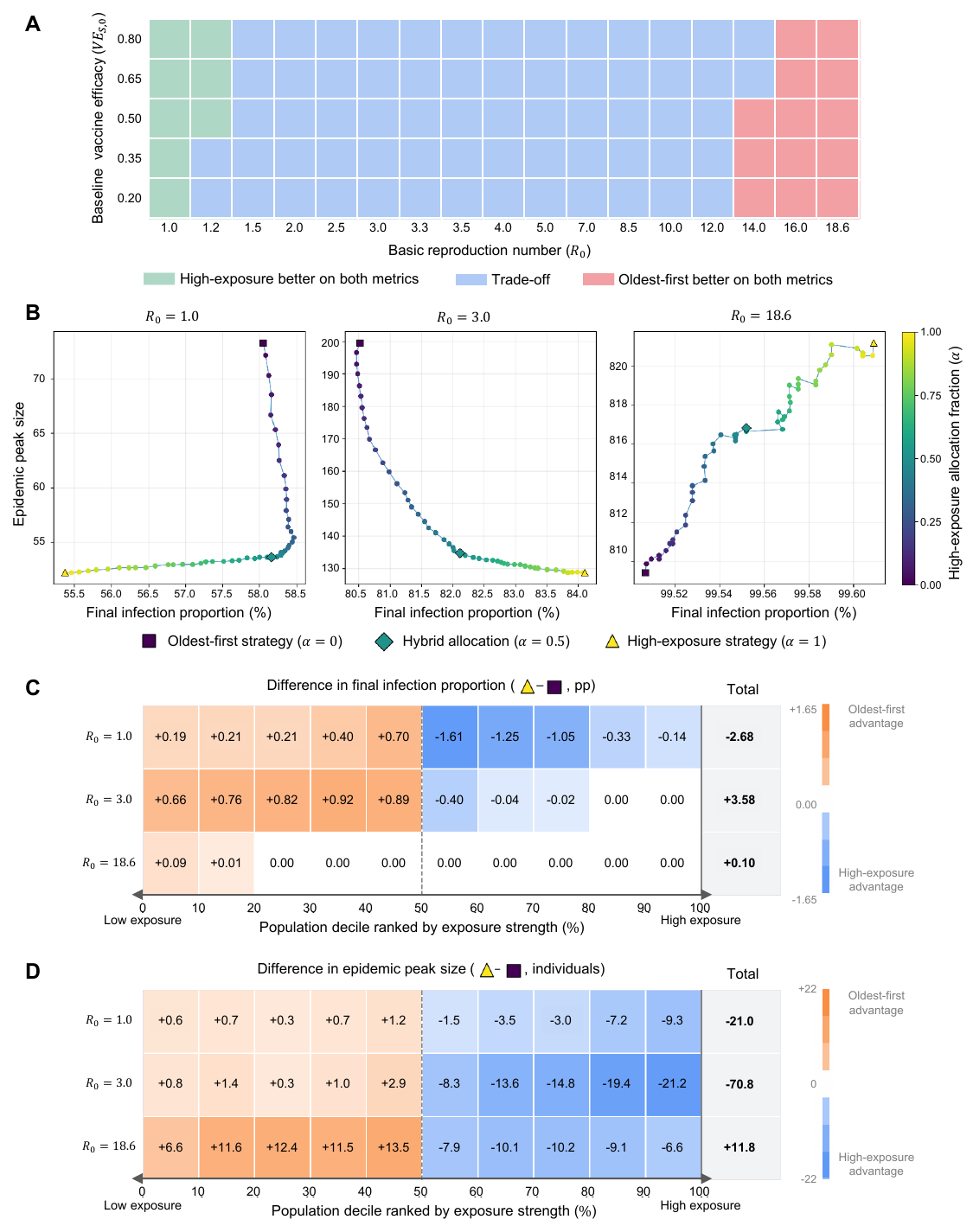}

    \captionsetup{font=small}

    \caption{Relative advantages of vaccination strategies across transmission intensities under leaky protection. All panels use $50\%$ vaccination coverage, with outcomes averaged across 100 Monte Carlo simulations per strategy or hybrid allocation. (\textit{A}) Comparison of the high-exposure and oldest-first strategies across combinations of $R_0$ and baseline vaccine efficacy $VE_{S,0}$. Green and red indicate lower values of both metrics under the high-exposure and oldest-first strategies, respectively. Blue indicates a trade-off in which the high-exposure strategy achieves a lower epidemic peak size at the cost of a higher final infection proportion than the oldest-first strategy. Regions are classified by endpoint performance, and panel \textit{B} examines the corresponding hybrid allocation trajectories in representative settings. (\textit{B}) Hybrid allocation trajectories at $VE_{S,0}=0.8$ for $R_0=1.0$, $3.0$ and $18.6$. Recipients selected by both targeted strategies remain vaccinated. The fraction $\alpha$ of the remaining vaccination places is assigned to the high-exposure strategy, with the remainder assigned to the oldest-first strategy. Points show mean outcomes at 51 equally spaced values of $\alpha$ from 0 to 1 and are coloured by $\alpha$. Squares, diamonds and triangles identify the oldest-first strategy ($\alpha=0$), hybrid allocation ($\alpha=0.5$) and high-exposure strategy ($\alpha=1$), respectively. (\textit{C}, \textit{D}) Exposure-decile contributions to the high-exposure minus oldest-first differences in final infection proportion (\textit{C}, percentage points) and epidemic peak size (\textit{D}, individuals) for the same scenarios as in \textit{B}. Residents are ranked by exposure strength from lowest to highest and divided into ten population deciles. In \textit{C}, differences in the number of individuals infected at least once in each group are divided by the total population and expressed in percentage points. In \textit{D}, infectious individuals in each group are counted at the peak in each simulation. Negative values (blue) and positive values (orange) indicate lower contributions under the high-exposure and oldest-first strategies, respectively. The dashed line separates the lower and higher exposure halves of the population. The Total column reports the overall difference between strategies. pp denotes percentage points.
    }
    \label{fig:hybrid_tradeoff}
\end{figure}

\clearpage

\subsection*{Vaccination strategies show three patterns of relative advantage as transmission intensity increases}

The preceding analyses show that the three epidemiological metrics do not always favour the same vaccination strategy. To further examine the role of exposure strength heterogeneity in these differences, we focus on epidemic peak size and final infection proportion. The metric of expected deaths is omitted from this comparison because its calculation includes additional weighting by age-specific IFRs. Based on the relative performance of the high-exposure and oldest-first strategies for these two metrics, we classify the parameter space into three transmission patterns (Fig.~\ref{fig:hybrid_tradeoff}{\textit A}). At low transmission intensity, the high-exposure strategy performs better on both metrics (green). At intermediate transmission intensity, the high-exposure strategy produces a lower epidemic peak size, whereas the oldest-first strategy produces a lower final infection proportion, indicating a trade-off (blue). At higher transmission intensity, the oldest-first strategy outperforms on both metrics (red).

To look into the underlying mechanism, we use representative scenarios in three patterns to examine how different hybrids of the two vaccination strategies affect final infection proportion and epidemic peak size (Fig.~\ref{fig:hybrid_tradeoff}\textit{B}). These hybrid allocations maintain a fixed vaccination coverage of $50\%$. We first retain all the overlap recipients that are selected by both strategies with such a vaccination coverage. Then, of the remaining vaccination slots, a fraction $\alpha$ is assigned to the high-exposure strategy and the remaining fraction $1-\alpha$ to the oldest-first strategy. Here, recipients are selected according to its original priority ranking related to the corresponding strategy. Note that the oldest-first strategy ($\alpha=0$) and the high-exposure strategy ($\alpha=1$) form the two endpoints. 

At $R_0=1.0$, the high-exposure endpoint outperforms the oldest-first endpoint on both metrics (left panel of Fig.~\ref{fig:hybrid_tradeoff}\textit{B}). As $\alpha$ increases, epidemic peak size decreases overall, whereas final infection proportion initially rises, reaches approximately 58.46\% at $\alpha=0.3$, and subsequently declines. In particular, the initial part of the hybrid allocation contains a local trade-off, in which peak suppression is accompanied by a slight increase in the population reach of infection. This turning point results from a progressive redistribution of vaccination across exposure levels. For $0< \alpha \le 0.3$, the additional vaccination places assigned to the high-exposure strategy are concentrated among individuals with the highest exposure strength, whose infection risk is already high. Prioritizing these individuals helps reduce epidemic peak size but may do little to prevent their eventual infection. Meanwhile, fewer vaccination places remain for lower-exposure individuals, which may contribute to the slight increase in final infection proportion relative to $\alpha=0$. As $\alpha$ increases further, protection extends to individuals with somewhat lower exposure strength, for whom infection is more preventable. Their exposure levels are low enough for vaccination to protect them effectively, but high enough for their protection to help limit transmission. This leads to simultaneous reductions in epidemic peak size and final infection proportion. 

This pattern changes for higher transmission intensity. At $R_0=3.0$, the two metrics change in opposite directions along the continuum: epidemic peak size decreases overall, whereas final infection proportion increases overall as $\alpha$ increases, producing a clear trade-off (middle panel of Fig.~\ref{fig:hybrid_tradeoff}\textit{B}). At $R_0=18.6$, both metrics increase overall with $\alpha$, and the oldest-first endpoint performs better on both. In this scenario, final infection proportion exceeds 99\% under every allocation, indicating that near-complete infection saturation strongly compresses the sensitivity of this outcome to different allocations (right panel of Fig.~\ref{fig:hybrid_tradeoff}\textit{B}). 

The distribution of endpoint differences across exposure levels helps explain these three patterns. We examine how each exposure decile contributes to the differences between the two endpoints in final infection proportion (Fig.~\ref{fig:hybrid_tradeoff}{\textit C}) and epidemic peak size (Fig.~\ref{fig:hybrid_tradeoff}{\textit D}). Individuals are ranked from lowest to highest exposure strength and divided into ten population deciles. Each table reports the endpoint difference, defined as the high-exposure endpoint minus the oldest-first endpoint. Negative values (blue) indicate that the high-exposure endpoint produces a lower value for the corresponding metric, which means it performs better than the oldest-first endpoint. Positive values (orange) indicate that the oldest-first endpoint performs better. Ten decile-specific differences in each row sum to the total value on the right. The advantage of the oldest-first endpoint is generally concentrated in the lower-exposure half, whereas the advantage of the high-exposure endpoint is concentrated in the higher-exposure half. Overall strategy performance is determined by the relative magnitudes of these opposing differences.

At $R_0=1.0$, for both metrics, the advantage of the high-exposure endpoint in the higher-exposure half outweighs the advantage of the oldest-first endpoint in the lower-exposure half. The total differences are therefore negative for both final infection proportion ($-2.68$ pp) and epidemic peak size ($-21.0$ individuals). At $R_0=3.0$, the balance differs between metrics. For final infection proportion, the modest advantage of the high-exposure endpoint in the higher-exposure half is insufficient to offset that of the oldest-first endpoint in the lower-exposure half, yielding a positive total difference ($3.58$ pp). For epidemic peak size, the high-exposure advantage in the higher-exposure half is substantially larger and exceeds the opposing difference in the lower-exposure half, yielding a negative total difference ($-70.8$ individuals). This separation is consistent with a broader range of exposure groups approaching infection as transmission intensity increases. Prioritizing individuals with higher exposure can still reduce the maximum concurrent infectious burden, but its benefit in preventing individuals from ever becoming infected is no longer sufficient to offset the loss of direct protection among lower-exposure groups. At $R_0=18.6$, differences in final infection proportion are close to zero across nearly all exposure deciles, indicating that infection saturation leaves little scope for either allocation to further alter the population reach of infection. The small residual difference is concentrated mainly in lower-exposure groups that retain some preventable infection risk. For epidemic peak size, the high-exposure endpoint still produces lower values in the higher-exposure half, but this advantage is smaller than the advantage of the oldest-first endpoint in the opposite half, making the total difference become positive ($11.8$ individuals). The endpoint reversal at very high transmission intensity does not imply that the high-exposure strategy completely loses its effect in higher-exposure groups. Rather, its localized advantage is no longer large enough to determine the population-level ordering of the metric.

Taken together, the three transmission patterns arise from changes in the balance of endpoint differences across the exposure distribution. The oldest-first endpoint generally performs better in the lower-exposure half, whereas the high-exposure endpoint generally performs better in the higher-exposure half. As transmission intensity increases, however, the relative magnitudes of these advantages change differently for the two epidemic metrics. This generates three qualitatively distinct patterns: high-exposure dominance for both metrics, a trade-off between the metrics, and oldest-first dominance for both metrics. We further find that these patterns depend on the mode of vaccine protection. We evaluate the two targeted strategies under all-or-nothing protection using the same three epidemiological metrics (\textit{SI Appendix}, Figs.~S1--S2). Endpoint rankings and hybrid allocation trajectories are shown in the \textit{SI Appendix}, Figs.~S3 and S4, respectively. Some high-transmission settings under all-or-nothing protection show opposite endpoint rankings for the two metrics, resembling those in the leaky trade-off region. However, in such all-or-nothing scenarios, increasing exposure prioritization reduces epidemic peak size while final infection proportion actually shows no clear trend. This indicates that opposite rankings of the two targeted strategies across the two metrics alone are insufficient to establish a consistent trade-off under hybrid allocation. The trajectories of both metrics as allocation varies must also be examined. 

\section*{Discussion}\label{sec3}

Vaccination is an important means of controlling infectious diseases, but its population-level impact depends on both vaccine protection mode and who is vaccinated. We examine how individual differences in exposure shape the relative performance of vaccination strategies under leaky protection. To do so, we use a survey-derived framework that combines an internal weighted contact network with external exposure and compare strategies across transmission intensities and epidemiological objectives. Our results show that, under leaky protection, the relative advantages of vaccination strategies depend on both the epidemiological objective and transmission intensity. The high-exposure strategy provides its most consistent benefit in reducing epidemic peak size, whereas the oldest-first strategy consistently reduces expected deaths relative to the random strategy. Importantly, the distinction between objectives extends beyond reducing infections versus reducing deaths. Epidemic peak size and final infection proportion can themselves favour different strategies, and increasing transmission intensity can reverse which strategy performs better on both metrics. These findings show that the value of exposure prioritization depends not only on whether it limits transmission, but also on which aspect of epidemic dynamics is being evaluated.

A possible explanation lies in the distinction between high infection risk and the opportunity to prevent infection through vaccination. Leaky protection reduces susceptibility at each exposure, but repeated infectious contacts can still produce a substantial cumulative probability of infection. Consequently, protection against individual exposures might not translate into equally strong protection over an epidemic~\cite{edlefsen2014leaky}. Vaccination may delay infection in highly exposed individuals without preventing it altogether, while allowing some individuals with lower exposure to remain uninfected. This provides a possible basis for reducing the simultaneous occurrence of infections without achieving a corresponding reduction in final infection proportion. It is also consistent with previous work showing that, under leaky protection, prioritizing lower-exposure groups can become advantageous for minimizing final epidemic size at sufficiently high transmission intensity~\cite{gavish2022optimal}. Our exposure-decile analysis adds a population-level interpretation. The overall strategy difference reflects the balance of gains and losses across exposure groups, rather than a uniform response throughout the population.

The all-or-nothing comparison helps clarify the role of the protection assumption. In the representative high-transmission all-or-nothing scenario, increasing exposure prioritization reduces epidemic peak size while final infection proportion varies over a narrow range (\textit{SI Appendix}, Fig. S4). This contrast highlights the distinction between opposite endpoint rankings and a sustained trade-off, which requires examining how the two metrics change together along hybrid allocation. Thus, vaccine failure mode should be treated as part of the explanation for strategy performance, rather than relying on vaccine efficacy alone~\cite{lee2025vaccine}. Note that the three transmission patterns identified here represent qualitative findings within the current modelling framework. Transitions between these patterns may occur at different $R_0$ values by altering transmission settings.

The present framework uses a survey-derived population and controlled comparisons to examine how leaky protection interacts with heterogeneous exposure. Its empirical basis in a single community provides a concrete setting for this analysis, while motivating further comparisons using independent community data. Such data would allow the effects of different joint distributions of age, household membership and activity to be assessed. More detailed contact observations and information on infection outside the community would also help constrain network reconstruction and external infection pressure. Moreover, our model already includes waning protection. Longitudinal information on vaccination, infection history and clinical outcomes would support extensions that distinguish more varied protection profiles and mortality risks. Together, these extensions would establish how broadly the identified relationships apply while retaining the controlled comparisons needed to interpret their mechanisms.

One implication for modelling studies is that epidemic peak size and final infection proportion should be evaluated jointly when comparing vaccination strategies. Reporting both metrics alongside the exposure distribution of vaccine recipients can help distinguish changes in the temporal concentration of infections from changes in the number of individuals infected. Hybrid allocation provides a useful way to examine this distinction while holding vaccination coverage fixed. It also reveals whether opposing rankings of two pure strategies correspond to a sustained trade-off as allocation changes. This offers a more informative assessment than either a single outcome or an endpoint comparison alone.

In summary, under leaky protection, transmission intensity changes both the relative advantages of vaccination strategies and the relationship between epidemic peak size and final infection proportion. High-exposure prioritization offers its most consistent benefit in peak reduction, while age-based prioritization reduces expected deaths under the assumed risk structure. Their effects on final infection proportion depend on how protection is distributed across heterogeneous exposure levels. These findings connect individual exposure differences with population-level outcomes and show why effective epidemic control cannot be represented by a single metric.

\section*{Materials and Methods}
\subsection*{Data source and construction of individual attributes}
We construct an individual-level contact and exposure network using data from a community-based household survey. The survey covered 1,034 households and 2,279 residents, and recorded demographic characteristics, household addresses, fixed (work or school activities) and non-fixed mobility behaviour, and behavioural features relevant to the transmission of acute respiratory infections. Each individual is represented as a node in the network, and information on age, mobility behaviour and household membership is converted into individual-level attributes.

To capture heterogeneity in exposure opportunities within and outside the community, we define two variables. First, reported work/school activity status is used together with the reported workplace or school population category to assign work/school-related external exposure in the model. Second, daily activity duration, which is truncated at 12 hours per day, combines the fixed work/school activity duration with the duration of non-fixed activities. The non-fixed activity duration is additionally used to construct stochastic community contacts (30\% of non-fixed duration) and the external exposure (70\% of non-fixed duration). Details of questionnaire coding, age-category mapping, variable standardization and data preprocessing are provided in the \textit{SI Appendix}, section~1A and Table~S1.

\subsection*{Ethics approval and informed consent}
The household survey was reviewed and approved by the Peking University Biomedical Ethics Committee. All data used in the present analysis were de-identified before analysis. Informed consent was obtained from all participants or from their parents or legal guardians, as applicable. All procedures were conducted in accordance with the relevant ethical guidelines and regulations.

\subsection*{Construction of the exposure framework}
We construct an individual-level exposure framework comprising an internal weighted contact network and individual-specific external exposure. The internal network contains household contacts representing stable interactions among co-residents and stochastic community contacts representing weaker activity-related interactions among residents. Both are encoded as weighted network edges. The external exposure represents exposure opportunities associated with work, school and non-fixed out-of-community activities and does not form network edges.

For each individual, exposure strength is defined as the sum of internal weighted network strength and the external exposure. The external exposure represents modelled exposure opportunities outside the internal contact network. External infection pressure additionally incorporates the transmission coefficient $\beta$ and external infectious prevalence. For both work/school and external non-fixed activities, external exposure is calculated by multiplying the assigned activity duration by the corresponding effective contact count. Internal weighted network strength and external exposure are both expressed in effective contact-hour equivalents per day. Network generation, edge-weight assignment and calculation of the external infection pressure are described in the \textit{SI Appendix}, sections~1B--1C and Table~S2.

\subsection*{Transmission dynamics}
We simulate transmission of acute respiratory infection on the weighted exposure network using a discrete-time, individual-based SEIS-type model with an exposed state and recovery-time-dependent partial immunity~\cite{hethcote2000mathematics,foxall2016seis,allen1994discrete}. At each daily time step, each individual is in one of three states: susceptible, exposed or infectious. Susceptible individuals enter the exposed state according to the combined hazard from infectious neighbours in the internal contact network and external infection pressure. Exposed individuals do not contribute to transmission and progress to the infectious state with daily probability $\kappa=0.2$. Infectious individuals recover with daily probability $\gamma=0.142$ and return to the susceptible state. These transitions are modelled as Bernoulli events at daily time steps, and the durations of the exposed and infectious states are geometrically distributed, with means $1/\kappa=5.0$ days and $1/\gamma\approx7.04$ days, respectively. We select these values as simplified COVID-like modelling assumptions. The exposed-state duration is chosen to approximate published estimates of the SARS-CoV-2 latent period, whereas estimates of infectious-period duration vary across empirical studies and transmission models~\cite{xin2022latent,byrne2020infectious}. Parameter settings for the influenza-like and RSV-like scenarios are provided in the \textit{SI Appendix}, section~1K and Tables~S4--S5. To allow reinfection while retaining partial protection from prior infection, individuals leaving the infectious state return to the susceptible state and are assigned a recovery timer; their subsequent susceptibility is reduced by a multiplier that depends on time since recovery.

We use values of the basic reproduction number \(R_0\) to define transmission intensities in the initially susceptible, unvaccinated population~\cite{diekmann1990definition,van2002reproduction,diekmann2010construction}. For each \(R_0\), we first multiply the target value by the internal exposure fraction, calculated from internal network exposure and prevalence-weighted external exposure. We then calibrate the transmission coefficient ($\beta$) using this adjusted value and the spectral radius of the internal weighted contact network~\cite{wang2003epidemic,van2008virus}. Details of this approximate calibration are provided in the \textit{SI Appendix}, section~1D and Table~S2. The specified \(R_0\) values do not represent the effective reproduction number after vaccination or infection-induced immunity changes the population susceptibility. 

For a susceptible individual \(i\), the total infection pressure on day \(t\) combines internal network transmission and external infection pressure:
\[
\lambda_i(t)=\frac{\beta}{24}\sum_j A_{ij}\mathbf{1}_{\{X_j(t)=I\}}+\lambda_i^{\mathrm{ext}},
\]
where \(A_{ij}\) denotes the weighted internal contact intensity between individuals \(i\) and \(j\), \(\mathbf{1}_{\{X_j(t)=I\}}\) indicates whether neighbour \(j\) in the internal contact network is infectious, and \(\lambda_i^{\mathrm{ext}}\) denotes external infection pressure.

The relative susceptibility of individual \(i\) on day \(t\), denoted by \(\sigma_i(t)\), incorporates both vaccine protection and post-infection natural immunity. Vaccine protection is modelled as leaky protection against susceptibility~\cite{smith1984assessment,halloran1992interpretation}. Vaccination reduces, but does not eliminate, the infection risk associated with each exposure. We denote the baseline vaccine efficacy by \(VE_{S,0}\). For vaccinated individuals, the vaccine-related susceptibility multiplier is \(\sigma_i^{\mathrm{vax}}(t)=1-VE_S(t)\), where \(VE_S(t)\) follows a stepwise waning profile. This multiplier is 1 for unvaccinated individuals. Post-infection natural immunity is represented by a time-dependent susceptibility multiplier, \(\sigma_i^{\mathrm{nat}}(t)\). The two multipliers act multiplicatively:
\[
\sigma_i(t)=\sigma_i^{\mathrm{vax}}(t)\sigma_i^{\mathrm{nat}}(t).
\]

Assuming that the infection hazard remains constant within each one-day interval and that successful infectious exposures occur independently, the probability that individual \(i\) escapes infection during day \(t\) is \(\exp[-\sigma_i(t)\lambda_i(t)]\). This expression follows standard stochastic epidemic formulations based on Poisson infectious contacts~\cite{britton2010stochastic}. The corresponding probability of at least one successful infection event is therefore
\[
p_i(t)=1-\exp[-\sigma_i(t)\lambda_i(t)].
\]
This complementary exponential form bounds the daily infection probability between 0 and 1 and allows infection pressure from multiple sources to accumulate on the hazard scale. Repeated exposure increases the cumulative probability of infection.

The vaccine-waning profile is informed by published evidence on protection against SARS-CoV-2 infection~\cite{goldberg2021waning,feikin2022duration}. The natural-immunity profile is informed by estimates of protection against Omicron reinfection~\cite{bobrovitz2023protective}. Exact breakpoints, multipliers and functional forms are provided in the \textit{SI Appendix}, section~1E and Table~S2.

\subsection*{Vaccination strategies}
We compare a random strategy with two main targeted vaccination strategies that have clear public health interpretations. The random strategy serves as the baseline which vaccinates individuals uniformly from the entire population. For this strategy, one vaccinated set is generated at each coverage level and reused across epidemiological parameter settings and Monte Carlo repetitions. The high-exposure strategy vaccinates individuals in descending order of exposure strength. It therefore prioritizes nodes with the largest sum of internal weighted network strength and the individual-specific external exposure, consistent with broader network-immunization principles~\cite{pastor2002immunization,cohen2003efficient}. This ranking reflects potential exposure opportunities rather than infection pressure and does not incorporate infectious prevalence or the infection states of other individuals. The oldest-first strategy vaccinates individuals in descending order of age, representing a vulnerability-oriented rule consistent with previous age-prioritization studies~\cite{medlock2009optimizing,bubar2021model}.

\subsection*{Epidemiological metrics}
We evaluate three epidemiological metrics. Final infection proportion is defined as the proportion of individuals infected at least once during the simulation. Epidemic peak size is the maximum number of individuals simultaneously infectious and serves as a measure of instantaneous epidemic burden. Expected deaths are estimated by summing age-specific IFRs over individuals infected at least once~\cite{verity2020estimates}. Since reinfection is allowed, individuals infected more than once contribute only once to final infection proportion and contribute their age-specific IFR only once to expected deaths. Expected deaths are measured as an IFR-weighted expectation over all the infected individuals rather than a count of simulated death events. The age-specific IFR values and exact calculation formula are provided in the \textit{SI Appendix}, Table~S5 and section~1G, respectively.

\subsection*{Strategy comparison}
To compare strategy performance, we calculate the additional reduction achieved by each strategy under the same $R_0$, $VE_{S,0}$ and vaccination coverage relative to the random strategy. For a metric $m$, let $Y_m^{0}$, $Y_m^{\mathrm{Random}}$ and $Y_m^s$ denote the mean outcomes across 100 Monte Carlo repetitions under no vaccination, the random strategy and targeted strategy $s$, respectively. The additional reduction is defined as
\begin{equation*}
\begin{split}
\Delta_m(s)
&=100\times\left[\frac{Y_m^{0}-Y_m^s}{Y_m^{0}}-\frac{Y_m^{0}-Y_m^{\mathrm{Random}}}{Y_m^{0}}\right]\\
&=100\times\frac{Y_m^{\mathrm{Random}}-Y_m^s}{Y_m^{0}}.
\end{split}
\end{equation*}
Thus, $\Delta_m(s)>0$ indicates that the targeted strategy achieves a greater outcome reduction than the random strategy. The result is expressed in percentage points. Monte Carlo uncertainty in the estimated outcomes and strategy differences is summarized in the \textit{SI Appendix}, Figs.~S9--S10.

\subsection*{Hybrid allocation}
For the hybrid allocation analysis, we first identify the recipient sets selected independently by the high-exposure and oldest-first strategies at a nominal vaccination coverage of $50\%$. In the reference network, each strategy selects 1,139 individuals, including 223 individuals selected by both strategies. These shared recipients remain vaccinated at every value of $\alpha$.

The remaining 916 vaccination places are divided between the two strategies. A fraction $\alpha$ is assigned to individuals selected only by the high-exposure strategy, and the remaining fraction $1-\alpha$ is assigned to individuals selected only by the oldest-first strategy, subject to integer rounding. Within each exclusive group, recipients are selected according to the corresponding strategy’s original priority ranking. Thus, $\alpha$ applies only to the remaining places, not to all vaccinated individuals.

We evaluate $\alpha$ at 51 equally spaced values from 0 to 1 in increments of 0.02. The allocations at $\alpha=0$ and $\alpha=1$ exactly reproduce the oldest-first and high-exposure recipient sets, respectively. Integer allocation procedures and further implementation details are provided in the \textit{SI Appendix}, section~1H.

\subsection*{Simulation protocol}
To systematically evaluate strategy performance across $R_0$ and \(VE_{S,0}\), we simulate a parameter space defined by these two parameters, along with vaccination coverage and vaccination strategy. $R_0$ spans low- to high-transmission scenarios, ranging from 1.0 to 18.6. $VE_{S,0}$ covers low to high efficacy, ranging from 0.20 to 0.80. Vaccination coverage ranges from 0 to 0.95 across 20 evenly spaced levels. For the vaccination-allocation experiments, each combination of $R_0$, $VE_{S,0}$, vaccination coverage and vaccination strategy is simulated for 500 days with 100 independent Monte Carlo repetitions, and 10 initial infectious individuals are selected at random in each simulation. The main text reports the additional reduction achieved by each targeted strategy relative to the random strategy at $50\%$ vaccination coverage. Comparisons of both targeted strategies across vaccination coverage levels are summarized in the \textit{SI Appendix}, Fig.~S5.

The exposure-decile decomposition uses the retained endpoint simulations from the hybrid allocation analysis at $R_0=1.0$, $3.0$ and $18.6$, with $VE_{S,0}=0.8$ and $50\%$ vaccination coverage (Fig.~\ref{fig:hybrid_tradeoff}\textit{C,D}). Individuals are ranked by exposure strength from lowest to highest and divided into ten population deciles, which remain the same for both strategies. For final infection proportion, we calculate the high-exposure minus oldest-first difference in the number of individuals infected at least once within each group. This difference is divided by the total population and expressed in percentage points. For epidemic peak size, we identify the population peak separately for each strategy in each simulation and count the infectious individuals in each group at that peak. The difference between these counts gives each group’s contribution to the overall peak-size difference.

Contributions are averaged across 100 paired Monte Carlo repetitions, with the same initial infectious individuals used for both strategies within each pair. For each metric, the contributions sum to the overall high-exposure minus oldest-first difference. Full calculation details are provided in the \textit{SI Appendix}, section~1H.

\section*{Author contributions}
Y.X. conceived the study, developed the model, performed the simulations and analyses, and drafted the manuscript. J.Z., H.C. and Z.Y. revised the manuscript. Q.W. and L.Q. contributed to the interpretation of the results. Z.W. and Z.J. provided funding support. All authors reviewed and approved the final manuscript.

\section*{Competing interests}
The authors declare no competing interest.

\section*{Data and code availability}
The epidemiological parameter values and their sources or modelling assumptions are described in the \textit{Materials and Methods} and the \textit{SI Appendix}, Tables~S2--S5. The household-survey data may be available from the corresponding author upon reasonable request. The custom code is available at \url{https://github.com/xyw980217/leaky-vaccine-network-model}.

\section*{Acknowledgments}
We thank all residents who participated in the household survey.

\bibliographystyle{unsrtnat}
\bibliography{leaky_vaccine}

\end{document}


\maketitle


\medskip

\section{Supplementary Methods}

\subsection{Survey preprocessing and exposure measures}
\label{sec:survey}

The household survey records demographic characteristics, household membership and daily activity patterns. Its summary lists 1,034 households. All 2,279 respondent records are retained. Column names and household identifiers are standardized, and the consolidated data are ordered by age category. Each resident corresponds to one node, with the consolidated row order supplying the node index.

Residents with the same community code and standardized household identifier are treated as members of the same household. The household identifier is recorded as an address. Address-format normalization affects 13 respondent records and reduces the number of distinct community--household combinations to 1,024. These 1,024 groups define the household contact layer; the difference from the survey's reported household count does not represent the exclusion of residents. Addresses identify household membership but do not determine distances between nodes.

Age categories are converted to the representative ages in Table~\ref{tab:survey_model_mapping}. Students and respondents working outside the home are assigned a daily work/school activity duration of 8 h. Category d denotes working from home or being unemployed and is not treated as evidence of regular attendance at a fixed workplace. Four respondents in this category nevertheless report workplace population counts. Review of these records identifies them as home-based freelance workers who also work outside the home without a fixed workplace. Their work-related exposure is retained using the same duration and contact-count mapping. The implementation includes category d in the 8-h duration coding to accommodate these exceptions. The other 230 respondents in category d report no workplace contacts, so their work-related external exposure is zero.

The questionnaire allows up to five non-fixed activity entries selected from ten venue categories. For each entry, the mapped weekly frequency is multiplied by the mapped duration and divided by seven. Their sum gives daily non-fixed activity duration. Venue labels do not determine whether an activity is assigned to internal or external exposure; all entries use the same 30\%--70\% split described below. Daily activity duration combines fixed and non-fixed activity and is bounded between 0 and 12 h. Community-contact and external-exposure calculations use the underlying non-fixed duration before this bound is applied.

The lowest frequency category, at most once per week, is mapped to 0.5 times per week, including 1,175 entries recorded with a full-width less-than sign. Three non-standard activity entries occur in two respondents: two report a frequency of fewer than two times per week and one reports one to three times per week; all three report a duration of less than 3 h. These duration responses do not match the categories used in the numerical mapping and are assigned 0 h. Consequently, these three entries add no time to the calculated daily non-fixed activity duration. The respondents and their other entries remain in the dataset. None of the age or workplace population responses requires an unrecognized-category fallback.

\begin{longtable}{L{0.18\textwidth}L{0.25\textwidth}L{0.23\textwidth}L{0.25\textwidth}}
\caption{\textbf{Survey variables and their numerical mappings.} Inputs come from the household survey. }\label{tab:survey_model_mapping}\\
\toprule Input & Categories or coding & Numerical mapping & Use in the model\\ \midrule
\endfirsthead
\multicolumn{4}{l}{\textit{Table~\ref{tab:survey_model_mapping} continued.}}\\
\toprule Input & Categories or coding & Numerical mapping & Use in the model\\ \midrule \endhead
\midrule \multicolumn{4}{r}{\textit{Continued on next page.}}\\ \endfoot
\bottomrule \endlastfoot
Age & a--g: $\leq18$, 19--29, 30--39, 40--49, 50--59, 60--69 and $\geq70$ years & 15, 24, 34.5, 44.5, 54.5, 64.5 and 75 years & Vaccination ranking and age-specific IFR assignment\\
Household membership & Community code and household identifier recorded as an address & One group per unique standardized combination & Household contact layer\\
Reported work/school activity status & a: student; b: working outside the home; c: retired; d: working from home or unemployed; e: other & Questionnaire categories retained; four reviewed category-d exceptions are described above & Work/school exposure assignment\\
Work/school activity duration & Categories a, b and d in the duration coding; categories c and e otherwise & 8 h per day for a, b and d; 0 h for c and e & Multiplied by the mapped contact count; the 230 category-d respondents reporting none have zero work-related exposure\\
Non-fixed activity entries & Up to five entries chosen from ten venue categories & Mapped frequency times duration, divided by seven, then summed over entries & Same internal/external split for every venue\\
Non-fixed activity frequency & None, at most once, 1--2, 3--4, or at least 5 times per week & 0, 0.5, 1.5, 3.5 or 5.5 times per week & Calculation of daily non-fixed activity duration\\
Non-fixed activity duration & None, up to 2 h, 2--5 h, or more than 5 h per occasion & 0, 1, 3.5 or 6 h & Calculation of daily non-fixed activity duration\\
Work/school population category & None; a: $\leq10$; b: 11--50; c: 51--99; d: $\geq100$ people & 0, 3, 10, 20 or 30 effective contacts & External exposure parameter adjusted to an effective contact count, not an observed number of contacts\\
Daily activity duration & Fixed plus non-fixed activity & Bounded to 0--12 h per day & Daily activity duration parameter\\
\end{longtable}

\subsection{Internal contacts and external exposure}
\label{sec:network}

The exposure framework combines an internal weighted contact network with individual external exposure. Household membership and individual attributes come from the survey. Community partners and external exposure weights are constructed from these attributes. Contact networks retain individual connections and contact heterogeneity~\cite{keeling2005networks,mossong2008social}, but the present network is a survey-derived construction rather than a complete record of observed contacts.

All members of the same model household are connected. We assume that each resident spends 12 h at home per day and divide this time equally among their household neighbours. Each household edge therefore receives a weight of 12 h divided by one less than the household size. For residents in households with at least two members, the household edge weights sum to 12 h. Single-person households have no household edges.

For community contacts generated from non-fixed activities, we assume one effective contact per 15 min, using a contact-duration unit also adopted in an indoor transmission model~\cite{li2022multinomial}. For community contacts, 30\% of each resident's daily non-fixed activity duration is allocated to the internal network. This duration is divided by 0.25 h and rounded to the nearest integer to determine how many partners the resident selects. Exact half-integers are rounded to the nearest even integer, as in the code. Partners are sampled uniformly without replacement from all other residents, up to the available population size. Each selection adds 0.25 to the corresponding undirected edge weight. The value 0.25 h is the duration of one effective contact on the model's contact-hour scale.

Partner sampling does not exclude household members. A selected household pair receives the community contribution in addition to its household weight. If both members of a pair select each other, both contributions are added to the same undirected edge. A resident's final community strength can therefore exceed the activity budget used for their own selections because other residents can also select them. The reference network is generated once with random seed 42 and remains fixed during the main simulations.

Let \(A\) be the symmetric weighted adjacency matrix of the internal network. Its entry \(A_{ij}\) is the total weight assigned to the connection between residents \(i\) and \(j\), and \(A_{ij}=0\) when no edge is present. There are no self-edges. The internal weighted network strength of resident \(i\) is \(\sum_j A_{ij}\).

External exposure is calculated separately for fixed work/school activities and non-fixed activities outside the internal network:
\[
V_i^{\mathrm{ext}}=V_i^{\mathrm{work}}+V_i^{\mathrm{social}}.
\]
For students, respondents working outside the home and the four reviewed freelance workers described in Section~\ref{sec:survey}, \(V_i^{\mathrm{work}}\) equals the mapped work/school contact count multiplied by 8 h. Residents with no assigned work/school activity or a zero contact count have \(V_i^{\mathrm{work}}=0\). This does not exclude external exposure from their non-fixed activities. The remaining $70\%$ of daily non-fixed activity duration is assigned to external activities. We use an effective contact count of 4 for these activities. This value is comparable in magnitude to the age-specific mean contact counts per activity used for adults in a previous transmission model~\cite{smieszek2011reconstructing}. Multiplying this count by the assigned duration gives \(V_i^{\mathrm{social}}\) in effective contact-hour equivalents per day, following the contact-hour approach of combining contact numbers with activity duration~\cite{glass2008social}. These external contacts are not represented by network edges.

The internal weights and external exposure are expressed on the same scale of effective contact-hour equivalents per day. These weights need not equal the time spent in distinct, non-overlapping encounters. For both work/school and external non-fixed activities, exposure is calculated by multiplying the assigned activity duration by the corresponding effective contact count. The exposure strength used for vaccination ranking is
\[
S_i=\sum_j A_{ij}+V_i^{\mathrm{ext}}.
\]
Neither \(\beta\) nor infectious prevalence enters this ranking. The external exposure \(V_i^{\mathrm{ext}}\) describes opportunities for exposure, irrespective of whether external contacts are infectious. External infection pressure \(\lambda_i^{\mathrm{ext}}\) is calculated from this quantity in Section~\ref{sec:dynamics}.

The reference internal network contains 2,279 nodes and 4,624 weighted edges. Its largest connected component contains 2,258 nodes and 4,617 edges and is shown in the main-text network visualization. All residents are retained in the transmission simulations and the age--exposure plot. Relative exposure strength is \(S_i/\max_j S_j\), with a population maximum of 268.5 in the reference network. This scaling does not change the vaccination ranking. Network positions are not geographical. Horizontal jitter within age groups reduces overlap in the age--exposure plot without changing exposure-strength values.

\subsection{Transmission dynamics and infection pressure}
\label{sec:dynamics}

We use the daily, individual-based SEIS-type model described in the \textit{Materials and Methods}. The state \(X_i(t)\) of resident \(i\) is susceptible (\(S\)), exposed (\(E\)) or infectious (\(I\)). Exposed individuals are infected but do not yet transmit infection. They progress to \(I\) with daily probability \(\kappa=0.20\). Infectious individuals recover with daily probability \(\gamma=0.142\) and return to \(S\). Their susceptibility after recovery depends on their recovery timer.

The waiting times in \(E\) and \(I\) follow geometric distributions with means \(1/\kappa=5.0\) days and \(1/\gamma\approx7.04\) days. These probabilities are discrete daily transition probabilities, not continuous-time transition rates. The 5-day mean is a simplified latent-period assumption informed by SARS-CoV-2 estimates, including a reported mean of 5.5 days~\cite{xin2022latent}. The infectious duration is a reference assumption informed by the variable estimates reviewed by Byrne et al.~\cite{byrne2020infectious}. Neither value is fitted to an outbreak in the surveyed community.

For a susceptible resident, the total infection pressure on day \(t\) is
\begin{equation}
\lambda_i(t)=\frac{\beta}{24}\sum_j A_{ij}\mathbf{1}_{\{X_j(t)=I\}}
+\lambda_i^{\mathrm{ext}},
\label{eq:total_pressure}
\end{equation}
where \(\mathbf{1}_{\{X_j(t)=I\}}\) identifies infectious neighbours in the internal contact network. External infection pressure is
\begin{equation}
\lambda_i^{\mathrm{ext}}=\frac{\beta}{24}V_i^{\mathrm{ext}}p_{\mathrm{ext}}.
\label{eq:external_pressure}
\end{equation}
Here, \(p_{\mathrm{ext}}=0.005\) is the assumed infectious prevalence outside the network. It is fixed throughout each simulation and does not respond to infections or vaccination among the modelled residents.

The coefficient \(\beta\) controls transmission on the daily exposure scale and is expressed per day. It is not a probability and need not lie between 0 and 1. Dividing the contact-hour weights by 24 normalizes them by the length of a day. Thus, \(\lambda_i(t)\) is an infection hazard, not an infection probability. With a one-day time step, its numerical value also gives the cumulative hazard during that day before susceptibility is applied. The individual probability of infection is calculated after combining this hazard with \(\sigma_i(t)\), as specified in Section~\ref{sec:immunity}.

\subsection{Calibration of the transmission coefficient}
\label{sec:calibration}

We use the specified \(R_0\) values to define transmission scenarios in the initially susceptible, unvaccinated population. The calibration incorporates both internal network exposure and external exposure. It first assigns a share of the target transmission scale to the internal network, then uses the weighted spectral radius to calculate \(\beta\).

The total internal weight is
\[
W_{\mathrm{int}}=\sum_i\sum_j A_{ij}.
\]
This is the sum of node strengths. Each undirected edge contributes twice, once for each endpoint. The external calibration weight is
\[
W_{\mathrm{ext}}=p_{\mathrm{ext}}\sum_i V_i^{\mathrm{ext}}.
\]
The external term is multiplied by infectious prevalence because only the assumed infectious fraction of external contacts contributes to infection pressure. The internal term uses all network weights because the reference calibration concerns transmission into an initially susceptible population from infectious network nodes. The internal exposure fraction used for calibration is
\begin{equation}
\pi=
\frac{W_{\mathrm{int}}}{W_{\mathrm{int}}+W_{\mathrm{ext}}}.
\label{eq:internal_fraction}
\end{equation}
It is an aggregate exposure adjustment used by the model, not the fraction of simulated infections attributed to the internal network.

Let \(\lambda_{\max}(A)\) denote the largest eigenvalue of the non-negative symmetric matrix \(A\), which equals its spectral radius. Under a linear approximation to internal transmission, the transmission-to-recovery scale is \(\beta\lambda_{\max}(A)/(24\gamma)\). The connection between network epidemic thresholds and the leading eigenvalue motivates this calibration~\cite{wang2003epidemic,van2008virus}. We assign the internal scale the value \(R_0\pi\), giving
\begin{equation}
\frac{\beta\lambda_{\max}(A)}{24\gamma}=R_0\pi,
\qquad
\beta=\frac{24\gamma R_0\pi}{\lambda_{\max}(A)}.
\label{eq:beta_calibration}
\end{equation}
The exposed-state probability \(\kappa\) controls the delay before infectiousness. It does not enter this calibration because every exposed individual eventually progresses to \(I\), and there is no competing removal from \(E\).

For the reference network, \(W_{\mathrm{int}}=26{,}574.5\), \(\sum_i V_i^{\mathrm{ext}}=150{,}853.5\), and \(W_{\mathrm{ext}}=754.2675\). The resulting \(\pi\approx0.972400\) and \(\lambda_{\max}(A)\approx12.848245\) give \(\beta\approx0.257929R_0\) when \(\gamma=0.142\). Across \(R_0=1.0\)--18.6, \(\beta\) therefore ranges from approximately 0.257929 to 4.797487 on the daily scale. The implementation adds \(10^{-9}\) to the denominator in Eq.~\ref{eq:internal_fraction} as a numerical safeguard.

The same calibrated \(\beta\) is used in Eqs.~\ref{eq:total_pressure} and \ref{eq:external_pressure}. Increasing \(R_0\) therefore increases both internal and external infection pressure, while \(p_{\mathrm{ext}}\) remains fixed. Calibration is repeated when the network or recovery probability changes, but not when vaccination coverage, strategy or \(VE_{S,0}\) changes. This procedure supplies a common transmission scale for the controlled comparisons. The exposure adjustment is a modelling approximation, not a next-generation calculation for the complete system with external introductions, waning and reinfection. The specified \(R_0\) is consequently not an estimate of the realized number of secondary infections in each simulation or the effective reproduction number after vaccination.

\subsection{Leaky protection and waning immunity}
\label{sec:immunity}

Under leaky protection, vaccination reduces susceptibility for every recipient~\cite{smith1984assessment,halloran1992interpretation}. The vaccine-related susceptibility multiplier is
\[
\sigma_i^{\mathrm{vax}}(t)=
\begin{cases}
1-VE_S(t), & \text{if individual }i\text{ is vaccinated},\\
1, & \text{otherwise}.
\end{cases}
\]
Baseline vaccine efficacy against susceptibility is denoted by \(VE_{S,0}\). All vaccination occurs before the simulation starts, and efficacy follows
\begin{equation}
VE_S(t)=
\begin{cases}
VE_{S,0}, & 1\leq t\leq100,\\
0.7VE_{S,0}, & 101\leq t\leq180,\\
0.4VE_{S,0}, & 181\leq t\leq500.
\end{cases}
\label{eq:vaccine_waning}
\end{equation}
Waning therefore depends on simulation day, not time since infection. Empirical evidence supports declining protection against SARS-CoV-2 infection~\cite{goldberg2021waning,feikin2022duration}. The two breakpoints and the relative efficacy values 1, 0.7 and 0.4 are a simplified model schedule, not a fitted vaccine-specific curve.

For a susceptible individual who has recovered, let \(\tau_i(t)\) be the recovery timer in days, which represents the days since the most recent recovery of individual $i$. The natural-immunity susceptibility multiplier is
\begin{equation}
\sigma_i^{\mathrm{nat}}(t)=
\begin{cases}
0.348, & 0\leq\tau_i(t)\leq90,\\
0.348+(0.753-0.348)\dfrac{\tau_i(t)-90}{275},
 & 90<\tau_i(t)\leq365,\\
0.753, & \tau_i(t)>365.
\end{cases}
\label{eq:natural_susceptibility}
\end{equation}
Before the first recovery, \(\sigma_i^{\mathrm{nat}}(t)=1\). The values 0.348 and 0.753 correspond to protection levels of 0.652 and 0.247. These protection values use the estimates at 3 and 12 months reported by Bobrovitz et al. as empirical anchors~\cite{bobrovitz2023protective}. Holding the first value constant through day 90, interpolating linearly to day 365 and retaining the residual value thereafter are modelling assumptions. The code resets the recovery timer at recovery, increments it during susceptible days and clears it upon reinfection. A newly recovered individual enters the next daily update with timer 1.

The two susceptibility multipliers act multiplicatively:
\[
\sigma_i(t)=\sigma_i^{\mathrm{vax}}(t)\sigma_i^{\mathrm{nat}}(t).
\]
Assuming a constant hazard within a day and independent infectious exposure events, the daily infection probability is
\begin{equation}
p_i(t)=1-\exp[-\sigma_i(t)\lambda_i(t)].
\label{eq:infection_probability}
\end{equation}
This expression follows the Poisson-contact formulation of infection risk~\cite{britton2010stochastic}. Vaccine efficacy reduces the hazard inside the exponential. It does not multiply the resulting daily probability by \(1-VE_S(t)\). For a specified exposure history, the probability of avoiding infection over successive susceptible days is the product of the daily escape probabilities, or \(\exp[-\sum_t\sigma_i(t)\lambda_i(t)]\). Repeated exposure can therefore increase cumulative infection risk despite partial protection.

Vaccination has no additional effect on infectiousness, progression, recovery or mortality conditional on infection. Reinfection is allowed, but final infection proportion and expected deaths count each infected resident only once.

\subsection{Parameter values and simulation design}
\label{sec:parameters}

Tables~\ref{tab:baseline_parameters} and \ref{tab:simulation_design} distinguish survey inputs, literature-informed values, derived quantities and model settings. A citation in a parameter row identifies the evidence informing that choice.

The main \(R_0\) grid spans weak to strong transmission rather than representing successive variants of one pathogen. Published estimates for early SARS-CoV-2 and Omicron illustrate the broad variation in transmission intensity across settings~\cite{liu2020reproductive,liu2022effective}. They provide context for the sweep, not empirical estimates for each grid point. In particular, 18.6 is the upper scenario value used here, not a value estimated from the community survey.

The five \(VE_{S,0}\) values span low to high partial protection. Variation in effectiveness against infection and its decline over time motivate examining a range of efficacy values~\cite{feikin2022duration}. The exact grid is a study design choice. Its entries are susceptibility reductions in the model and are not equated directly with observational vaccine-effectiveness estimates. The same disease-course parameters are retained across the main grid to isolate differences associated with transmission intensity, efficacy and vaccination strategy.

\begin{longtable}{L{0.25\textwidth}L{0.20\textwidth}L{0.17\textwidth}L{0.29\textwidth}}
\caption{\textbf{Reference model parameters, derived quantities and their basis.} Contact weights are effective contact-hour equivalents per day. Susceptibility multipliers, efficacy and prevalence are dimensionless. Disease progression and recovery are probabilities per daily update. Model settings are held fixed in the principal comparisons rather than estimated from infection data.}\label{tab:baseline_parameters}\\
\toprule Quantity & Symbol or definition & Reference value & Source and interpretation\\ \midrule
\endfirsthead
\multicolumn{4}{l}{\textit{Table~\ref{tab:baseline_parameters} continued.}}\\
\toprule Quantity & Symbol or definition & Reference value & Source and interpretation\\ \midrule \endhead
\midrule \multicolumn{4}{r}{\textit{Continued on next page.}}\\ \endfoot
\bottomrule \endlastfoot
Population & \(N\) & 2,279 & Household survey.\\
Model household groups & Cleaned community--address combinations & 1,024 & Derived from survey identifiers.\\
Daily time at home & Duration allocated to household contacts & 12 h & Daily home duration also used by Hartnett et al.~\cite{hartnett2021modelling}. Here, it is shared among household neighbours when the household has at least two residents.\\
Work/school activity duration & Daily duration assigned from activity status & 8 h for categories a, b and d; 0 otherwise & An 8 h daytime activity period is also used in contact-network modelling~\cite{hartnett2021modelling}. Category-d exceptions and zero contact counts are explained in Section~\ref{sec:survey}.\\
Effective work/school contact count & Workplace or school population mapped to effective contacts & 0, 3, 10, 20, 30 & Mapping in Table~\ref{tab:survey_model_mapping}. Contact counts need not increase in proportion to venue population, as represented in previous contact-network models~\cite{moreno2021anatomy}.\\
Within-community fraction of non-fixed activity & Fraction of non-fixed activity duration assigned to the internal network & 0.30 & A 30\% within-community allocation is used here. Partitioning activity time between internal and external transmission is also used by Dewald et al.~\cite{dewald2022effective}.\\
Effective community contact duration & Duration assigned to each community contact & 0.25 h (15 min) & Used to calculate the number of community partners selected from the allocated internal activity duration and the weight added per selection. A 15 min contact unit is used in an indoor transmission model~\cite{li2022multinomial}.\\
Outside-community fraction of non-fixed activity & Fraction assigned to external exposure & 0.70 & Complement of the within-community fraction, using the same time partition.\\
Effective contact count for external non-fixed activities & Effective contact count used to calculate \(V_i^{\mathrm{social}}\) & 4 & Multiplied by the daily non-fixed activity duration allocated outside the community to obtain \(V_i^{\mathrm{social}}\) in effective contact-hour equivalents per day.\\
External infectious prevalence & \(p_{\mathrm{ext}}\) & 0.005 & A common background prevalence of 0.5\% is used here. The external infection pressure depends on prevalence and transmissibility, as in Dewald et al.~\cite{dewald2022effective}.\\
Internal weighted contact intensity & \(A_{ij}\) & Constructed per edge & Household and community contributions in Section~\ref{sec:network}; edge weights represent contact intensity in a weighted network~\cite{barrat2004architecture}.\\
External exposure & \(V_i^{\mathrm{ext}}\) & Varies by resident & Sum of work/school and external non-fixed activity exposure. Each component is calculated by multiplying its assigned activity duration by the corresponding effective contact count.\\
Exposure strength & \(S_i\) & \(\sum_j A_{ij}+V_i^{\mathrm{ext}}\) & Extends weighted node strength~\cite{barrat2004architecture} by adding external exposure. Measures individual exposure and determines vaccination ranking, independently of infectious prevalence and \(\beta\).\\
Internal exposure fraction & \(\pi\) & 0.972400 & Calculated for the reference network using Eq.~\ref{eq:internal_fraction}; used to adjust the target \(R_0\) before spectral calibration.\\
Weighted spectral radius & \(\lambda_{\max}(A)\) & 12.848245 & Largest eigenvalue of the reference adjacency matrix, used for network-based transmission calibration~\cite{wang2003epidemic,van2008virus}.\\
Transmission coefficient & \(\beta\), per day & \(0.257929R_0\), approximately & Derived, Eq.~\ref{eq:beta_calibration}. Spectral calibration~\cite{wang2003epidemic,van2008virus}, with the model's external-exposure adjustment.\\
External infection pressure & \(\lambda_i^{\mathrm{ext}}\) & \(\beta V_i^{\mathrm{ext}}p_{\mathrm{ext}}/24\) & Daily external infection hazard from individual exposure, transmissibility and external infectious prevalence~\cite{dewald2022effective}; not an infection probability.\\
Daily progression probability & \(\kappa\) & 0.20 & Corresponds to a mean latent duration of 5 days, informed by latent-period estimates~\cite{xin2022latent}.\\
Daily recovery probability & \(\gamma\) & 0.142 & Corresponds to a mean infectious duration of 7.04 days, informed by a review of infectious-period estimates~\cite{byrne2020infectious}.\\
Vaccine efficacy waning & \(VE_S(t)/VE_{S,0}\) & 1, 0.7, 0.4 & Piecewise representation with breakpoints at days 100 and 180, informed by evidence of declining vaccine protection~\cite{goldberg2021waning,feikin2022duration}.\\
Vaccine-related susceptibility multiplier & \(\sigma_i^{\mathrm{vax}}(t)\) & \(1-VE_S(t)\) if vaccinated, otherwise 1 & Remaining susceptibility under leaky protection~\cite{smith1984assessment,halloran1992interpretation}.\\
Natural-immunity susceptibility multiplier & \(\sigma_i^{\mathrm{nat}}(t)\) & 0.348 initially, 0.753 after day 365 & Complements of the protection estimates 0.652 and 0.247~\cite{bobrovitz2023protective}.\\
Time since recovery & \(\tau_i(t)\) & Days since the most recent recovery & Determines natural-immunity susceptibility through Eq.~\ref{eq:natural_susceptibility}.\\
Relative susceptibility & \(\sigma_i(t)\) & \(\sigma_i^{\mathrm{vax}}(t)\sigma_i^{\mathrm{nat}}(t)\) & Product of remaining susceptibility after vaccination and infection. Multiplicative protection is also used in a model of hybrid immunity~\cite{larsen2025influence}.\\
Age-specific infection fatality ratio & \(\mathrm{IFR}_i\) & Table~\ref{tab:ifr_profiles} & COVID-like age-specific IFR profile~\cite{verity2020estimates}.\\
\end{longtable}

\begin{longtable}{L{0.22\textwidth}L{0.30\textwidth}L{0.39\textwidth}}
\caption{\textbf{Parameter grids and numerical settings.} Exact grid points, coverage levels and run sizes are experimental design choices. Literature citations provide context for the selected range rather than direct estimates of every value.}\label{tab:simulation_design}\\
\toprule Quantity & Values & Basis and purpose\\ \midrule
\endfirsthead
\multicolumn{3}{l}{\textit{Table~\ref{tab:simulation_design} continued.}}\\
\toprule Quantity & Values & Basis and purpose\\ \midrule \endhead
\midrule \multicolumn{3}{r}{\textit{Continued on next page.}}\\ \endfoot
\bottomrule \endlastfoot
Transmission intensity, \(R_0\) & 1, 1.2, 1.5, 2, 2.5, 3, 3.3, 3.5, 4, 5, 7, 8.5, 10, 12, 14, 16, 18.6 & Broad transmission sweep. Published estimates motivate examining varied intensities~\cite{liu2020reproductive,liu2022effective}.\\
Baseline efficacy, \(VE_{S,0}\) & 0.20, 0.35, 0.50, 0.65, 0.80 & Five levels of partial protection, giving 85 \(R_0\)--\(VE_{S,0}\) combinations. Effectiveness varies across settings and time~\cite{feikin2022duration}.\\
Vaccination coverage, \(c\) & 0--0.95 in increments of 0.05 & Resource-constraint sensitivity. Zero coverage supplies the no-vaccination reference.\\
Main comparison coverage & 0.50 & Same recipient count under all three strategies, \(\lfloor0.5N\rfloor=1{,}139\).\\
Simulation horizon & 500 days & Common observation period across strategies, including the specified waning stages.\\
Time step & 1 day & Synchronous daily updates with Bernoulli state transitions within a stochastic epidemic framework~\cite{britton2010stochastic}.\\
Initial infectious individuals & 10, sampled uniformly without replacement & Common initialization rule across strategies, not fitted to observed introductions.\\
Monte Carlo repetitions & 100 per strategy or hybrid allocation & Precision is assessed using Monte Carlo standard errors~\cite{morris2019simulation}; see Section~\ref{sec:uncertainty}.\\
Hybrid fraction, \(\alpha\) & 0--1 in increments of 0.02 & 51 allocations of the 916 vaccination places not shared by the endpoints.\\
Main hybrid scenarios & \(R_0=1.0,3.0,18.6\), \(VE_{S,0}=0.80\), \(c=0.50\) & Representative scenarios for the main-text trajectories and decile decomposition.\\
Matched leaky--all-or-nothing scenarios & \(R_0=1.0,7.0,18.6\), \(VE_{S,0}=0.80\), \(c=0.50\) & Comparisons under matched transmission, efficacy and coverage settings.\\
Reference random seed & 42 for network construction and principal scenario workers & Reproducible numerical setting.\\
\end{longtable}

Each replicate starts with 10 uniformly sampled infectious individuals. All others start susceptible without natural immunity. Initial infection is imposed independently of vaccination status. States update synchronously for 500 days using the states at the start of each day. Newly exposed individuals cannot progress to infectiousness in the same update, and newly recovered individuals cannot be reinfected until a later update. Daily infectious counts are recorded after each update.

The main parameter sweep resets its pseudorandom generators to seed 42 at the start of each \(R_0\)--\(VE_{S,0}\) worker. Replicates within a setting use successive draws. Separate parameter combinations are therefore reproducible but are not assigned independent master streams. The paired design used for hybrid allocation is described separately in Section~\ref{sec:hybrid}.

\subsection{Vaccination strategies and epidemiological metrics}
\label{sec:strategies}

At coverage \(c\), the three strategies vaccinate \(\lfloor cN\rfloor\) individuals. The random strategy selects recipients uniformly without replacement. The high-exposure strategy selects nodes in descending order of \(S_i\). The oldest-first strategy selects nodes in descending order of representative age. The implemented sorting order resolves tied values. In the main parameter sweep, each strategy's recipient set at a given coverage is reused across epidemiological parameter settings and Monte Carlo repetitions. No doses are reallocated during an epidemic.

For replicate \(r\), let \(J_{ir}\) indicate whether node \(i\) is infected at least once, including initial seeds. The three metrics are
\[
F_r=\frac{1}{N}\sum_iJ_{ir},\qquad
P_r=\max_{1\leq t\leq500}\sum_i\mathbf{1}\{X_{ir}(t)=I\},
\qquad
D_r=\sum_iJ_{ir}\,\mathrm{IFR}_i.
\]
\(F_r\) is final infection proportion, \(P_r\) is epidemic peak size and \(D_r\) is expected deaths. The IFR mapping is given in Table~\ref{tab:ifr_profiles}. For an age outside the mapped values, the nearest representative age is used. Deaths are not simulated as a separate state and do not remove nodes from transmission. This metric is an IFR-weighted estimate based on unique infected individuals, not a count of observed or simulated fatal events.

For each metric \(m\), the quantities \(Y_m^0\), \(Y_m^{\mathrm{Random}}\) and \(Y_m^s\) are the mean outcomes across 100 Monte Carlo repetitions under no vaccination, the random strategy and targeted strategy \(s\), respectively. Comparisons hold \(R_0\), \(VE_{S,0}\) and vaccination coverage fixed. The shared no-vaccination mean comes from the random-strategy simulations at zero coverage in the matching parameter setting. Additional reduction is
\[
\Delta_m(s)=100\frac{Y_m^{\mathrm{Random}}-Y_m^s}{Y_m^0}.
\]
This is the difference between two percentage reductions relative to no vaccination. Its unit is percentage points (pp), including when the underlying metric is a count. Positive values favour the targeted strategy. It is distinct from the percentage reduction relative to the random-strategy outcome.

\subsection{Hybrid allocation and endpoint decomposition}
\label{sec:hybrid}

Hybrid allocation varies the contribution of the two targeted strategies while holding nominal vaccination coverage at 50\%. In the reference network, each strategy selects 1,139 residents, including 223 common recipients and 916 recipients exclusive to each strategy. Let \(H\) and \(O\) be the recipient sets selected by the pure high-exposure and oldest-first strategies. Their common recipients, \(C=H\cap O\), remain vaccinated at every value of \(\alpha\). Each exclusive set, \(H\setminus C\) or \(O\setminus C\), contains \(M=\lfloor0.5N\rfloor-|C|\) individuals. We assign
\[
n_H(\alpha)=\left\lfloor\alpha M+0.5\right\rfloor,
\qquad n_O(\alpha)=M-n_H(\alpha)
\]
of the remaining vaccination places to these two exclusive sets. Within each set, selection follows the corresponding original priority ranking. Thus, \(\alpha\) specifies the share of the remaining vaccination places assigned to the high-exposure strategy, subject to integer rounding. The endpoints \(\alpha=0\) and \(\alpha=1\) reproduce the pure oldest-first and high-exposure strategies exactly.

We evaluate 51 equally spaced values of \(\alpha\) from 0 to 1 in increments of 0.02 at \(VE_{S,0}=0.80\). The main representative scenarios use \(R_0=1.0,3.0\) and 18.6. The matched leaky--all-or-nothing comparison uses \(R_0=1.0,7.0\) and 18.6. Each point represents the mean of 100 Monte Carlo simulations. Within a model and parameter setting, the same replicate index uses the same initial infectious seeds and daily random draws across \(\alpha\). This common-random-number design permits paired differences along the hybrid allocation. It does not imply pairing between leaky and all-or-nothing simulations, or with the separate full-grid experiments.

The endpoint map is calculated directly from the retained full-grid means used for the strategy heatmaps. Define
\[
\delta_F=\overline{F}_{H}-\overline{F}_{O},
\qquad
\delta_P=\overline{P}_{H}-\overline{P}_{O}.
\]
Negative values for both metrics indicate a high-exposure advantage. Positive values for both indicate an oldest-first advantage. Opposite signs indicate different endpoint rankings for the two metrics. This classification uses unrounded means. It does not test significance, impose a practical equivalence margin or establish a monotonic relationship along \(\alpha\).

For the main endpoint decomposition, residents are ranked by exposure strength from lowest to highest and divided into ten population deciles, with node index resolving ties. Nine groups contain 228 nodes and one contains 227. Groups remain fixed across strategies. For group \(g\), its contribution to the final infection proportion difference is
\[
d_{F,g,r}=\frac{100}{N}\sum_{i\in g}(J_{ir}^{H}-J_{ir}^{O}).
\]
For epidemic peak size, we count infectious individuals in each group at the population peak under each strategy in each replicate. If \(t_{H,r}^{*}\) and \(t_{O,r}^{*}\) are the first days attaining the respective maxima, the contribution is
\[
d_{P,g,r}=\sum_{i\in g}\mathbf{1}\{X_{ir}^{H}(t_{H,r}^{*})=I\}
-\sum_{i\in g}\mathbf{1}\{X_{ir}^{O}(t_{O,r}^{*})=I\}.
\]
The two strategies can reach their peaks on different days. Contributions sum to the total endpoint difference within every replicate. We average the paired contributions and calculate their Monte Carlo standard errors from the replicate differences. The lower and upper five deciles can be summed within each replicate to obtain contributions from the two exposure halves.

\subsection{All-or-nothing comparison with matched waning}
\label{sec:aon}

The all-or-nothing (AON) comparison changes how vaccine protection is distributed among recipients. All network, transmission, natural-immunity, vaccination-ranking and outcome settings remain as above. Unlike the leaky model, an all-or-nothing recipient is either fully protected against acquisition or receives no vaccine protection.

Let \(v_i=1\) for a vaccinated resident and \(v_i=0\) otherwise. For each node and Monte Carlo replicate, we draw \(u_i\sim\mathrm{Uniform}(0,1)\). The vaccine-protected indicator is
\[
q_i(t)=v_i\mathbf{1}\{u_i<VE_S(t)\}.
\]
The expected protected fraction among recipients is therefore \(VE_S(t)\). The realized number varies between replicates rather than being fixed to an exact proportion. For this comparison, the vaccine susceptibility multiplier is \(\sigma_i^{\mathrm{vax}}(t)=1-q_i(t)\), while \(\sigma_i^{\mathrm{nat}}(t)\) remains as in Eq.~\ref{eq:natural_susceptibility}. The all-or-nothing infection probability is
\[
p_i^{\mathrm{AON}}(t)=
1-\exp\!\left[-(1-q_i(t))\sigma_i^{\mathrm{nat}}(t)\lambda_i(t)\right].
\]
Protected susceptible nodes have zero acquisition probability. Nonresponders and recipients whose protection has waned follow the same susceptibility rule as unvaccinated nodes with the same natural-immunity status.

The same \(u_i\) is retained through all waning stages. Protected sets are nested as their expected proportions decrease from \(VE_{S,0}\) to \(0.7VE_{S,0}\) and then \(0.4VE_{S,0}\). Lost protection is not reassigned to other recipients. The uniform draws use a separate reproducible random stream and are shared at node level across strategies within the corresponding setting.

The original seeding convention is retained. A recipient who would otherwise be protected can still be selected as one of the 10 initial infectious seeds and is assigned infection at day 0. These infections are imposed at initialization and are not reversed by all-or-nothing protection. Subsequent acquisition follows the protected indicator above.

The all-or-nothing heatmaps use 100 replicates for each of the same 85 parameter combinations at 50\% coverage. Their endpoint map uses those same retained means (Fig.~\ref{fig:aon_endpoint}). The matched hybrid comparison uses the three scenarios specified in Section~\ref{sec:hybrid}. Within each all-or-nothing scenario, the protection draws and epidemic random streams are shared across \(\alpha\). We do not pool these runs with the separate full-grid runs.

\subsection{Sensitivity to vaccination coverage}
\label{sec:robustness}

Coverage sensitivity uses the retained leaky simulations from 5\% to 95\% coverage in 5 pp increments. At every coverage, each targeted strategy is compared with the random strategy at that same coverage. Means and interquartile ranges across the 85 parameter settings summarize the results. These ranges represent variation between scenarios, not Monte Carlo uncertainty.

\subsection{Sensitivity to disease parameters}
\label{sec:disease_robustness}

The disease sensitivity analysis uses COVID-like, influenza-like and RSV-like parameter sets (Tables~\ref{tab:disease_parameters} and \ref{tab:ifr_profiles}). Each profile uses its own reference \(R_0\), respectively 2.6, 1.28 and 1.7, and \(VE_{S,0}=0.20,0.50,0.80\). All nine settings retain the same network, 50\% coverage, 10 initial seeds, 500 simulation days and 100 replicates. Disease course, immunity timing and IFR weights change together. These profiles are sensitivity scenarios, not calibrated forecasts for three pathogens.

Influenza and RSV incubation estimates inform the exposed-state durations~\cite{lessler2009incubation}. Using them as latent-period proxies is a modelling assumption. The influenza infectious-period proxy uses a mean viral-shedding duration of 4.8 days~\cite{carrat2008influenza}. Its reference \(R_0\) is informed by published seasonal-influenza estimates~\cite{biggerstaff2014estimates}. Evidence of vaccine waning informs the use of time-dependent protection for the COVID-like, influenza-like and RSV-like scenarios~\cite{goldberg2021waning,feikin2022duration,young2018duration,ison2024rsv}. The stepwise schedules use selected breakpoints and relative efficacy levels to compare different protection durations. In particular, the RSV-like 90- and 180-day breakpoints define a short-duration sensitivity profile rather than a fit to the RSV vaccine trial, which measures protection against respiratory illness over two seasons~\cite{ison2024rsv}. For natural immunity, the COVID-like profile uses protection estimates at 3 and 12 months~\cite{bobrovitz2023protective}. Influenza and RSV reinfection studies support incomplete protection after infection~\cite{memoli2020influenza,hall1991rsv}. The corresponding plateau lengths, subsequent decline and residual susceptibility values are selected sensitivity settings.

The baseline IFR profile retains the main COVID-like mapping~\cite{verity2020estimates}. The influenza profile adapts the symptomatic case-fatality estimates reported by McDonald et al.~\cite{mcdonald2023inference}. For representative ages 15 and 24, it uses the 15--19 and 20--24 estimates. For ages 34.5, 44.5, 54.5 and 64.5, it takes the arithmetic means of the two corresponding five-year age bands. The oldest category uses the 75--79 estimate. This gives symptomatic case-fatality values of 0.001\%, 0.002\%, 0.0035\%, 0.007\%, 0.018\%, 0.0745\% and 0.49\%, respectively. Multiplying their probability equivalents by the symptomatic fraction of 0.669 gives the influenza-like IFR proxy~\cite{carrat2008influenza}. The unweighted age-band averaging and use of one symptomatic fraction across ages are model approximations.

The RSV profile represents the higher adult mortality burden at older ages~\cite{hansen2022mortality}. For adults aged 60 years and older, Savic et al. report pooled annual attack rates of 1.62\% for RSV-associated acute respiratory illness and 0.15\% for hospitalization, with a hospital case-fatality ratio of 7.13\%~\cite{savic2023burden}. Combining these estimates gives \(0.0015\times0.0713/0.0162\approx0.0066\), which provides a scale for the risk weight at representative age 64.5. This calculation relates hospital deaths to symptomatic illness, not all infections. The remaining RSV weights specify an age-dependent sensitivity profile rather than directly measured age-specific IFRs. The survey's youngest category is represented by age 15, so the profile does not resolve infant RSV risk.

\begin{longtable}{L{0.23\textwidth}L{0.12\textwidth}L{0.12\textwidth}L{0.12\textwidth}L{0.30\textwidth}}
\caption{\textbf{Disease-course and immunity parameters in the sensitivity analysis.} Transition probabilities apply to one-day updates. Durations are the means of geometric waiting times. Natural-immunity entries are remaining susceptibility, \(\sigma_i^{\mathrm{nat}}\), not protective efficacy. Each profile uses the same piecewise forms as Eqs.~\ref{eq:vaccine_waning} and \ref{eq:natural_susceptibility}, with its own values and interval lengths.}\label{tab:disease_parameters}\\
\toprule Parameter & COVID-like & Influenza-like & RSV-like & Basis\\ \midrule
\endfirsthead
\multicolumn{5}{l}{\textit{Table~\ref{tab:disease_parameters} continued.}}\\
\toprule Parameter & COVID-like & Influenza-like & RSV-like & Basis\\ \midrule \endhead
\midrule \multicolumn{5}{r}{\textit{Continued on next page.}}\\ \endfoot
\bottomrule \endlastfoot
Reference \(R_0\) & 2.60 & 1.28 & 1.70 & Selected reference scenarios. COVID-like value lies within early estimates~\cite{liu2020reproductive}. Influenza value follows the seasonal median~\cite{biggerstaff2014estimates}. RSV value is a low-transmission scenario within the model-dependent range discussed by Reis and Shaman~\cite{reis2016retrospective}.\\
Progression probability, \(\kappa\) & 0.200000 & 0.714286 & 0.227273 & Daily progression probabilities corresponding to the exposed-state durations below, informed by latent- and incubation-period estimates~\cite{xin2022latent,lessler2009incubation}.\\
Recovery probability, \(\gamma\) & 0.142000 & 0.208333 & 0.200000 & Daily recovery probabilities corresponding to mean infectious durations of 7.04, 4.80 and 5.00 days, respectively~\cite{byrne2020infectious,carrat2008influenza,reis2016retrospective}.\\
Mean exposed duration, days & 5.00 & 1.40 & 4.40 & COVID-like assumption~\cite{xin2022latent}. Influenza and RSV incubation estimates are used as latent-period proxies~\cite{lessler2009incubation}.\\
Mean infectious duration, days & 7.04 & 4.80 & 5.00 & COVID-like reference~\cite{byrne2020infectious}. Influenza uses mean shedding duration~\cite{carrat2008influenza}. RSV uses a short-duration assumption within previously modelled ranges~\cite{reis2016retrospective}.\\
End of initial vaccine stage, day & 100 & 60 & 90 & Selected first breakpoints in stepwise waning profiles, informed by evidence of declining vaccine protection~\cite{feikin2022duration,young2018duration,ison2024rsv}.\\
End of intermediate vaccine stage, day & 180 & 120 & 180 & Selected second breakpoints in the same profiles~\cite{feikin2022duration,young2018duration,ison2024rsv}. See Section~\ref{sec:disease_robustness} for the protection profiles.\\
\(VE_S(t)/VE_{S,0}\) across stages & 1, 0.7, 0.4 & 1, 0.7, 0.4 & 1, 0.7, 0.4 & Shared sensitivity levels for a stepwise approximation of waning protection~\cite{feikin2022duration,young2018duration,ison2024rsv}.\\
Initial \(\sigma_i^{\mathrm{nat}}\) after recovery & 0.348 & 0.500 & 0.600 & COVID-like complement of 0.652 protection~\cite{bobrovitz2023protective}. Influenza-like and RSV-like values are sensitivity levels motivated by incomplete protection against reinfection~\cite{memoli2020influenza,hall1991rsv}.\\
Late \(\sigma_i^{\mathrm{nat}}\) & 0.753 & 0.900 & 0.950 & COVID-like complement of 0.247 protection~\cite{bobrovitz2023protective}. Influenza-like and RSV-like values specify residual protection in the reinfection sensitivity profiles~\cite{memoli2020influenza,hall1991rsv}.\\
End of natural-immunity plateau, day & 90 & 60 & 60 & COVID-like 3-month evidence anchor~\cite{bobrovitz2023protective}. The other profiles use a selected early interval in light of incomplete protection after infection~\cite{memoli2020influenza,hall1991rsv}.\\
End of linear susceptibility increase, day & 365 & 180 & 180 & COVID-like 12-month evidence anchor~\cite{bobrovitz2023protective}. The other profiles use a selected later interval for waning natural protection~\cite{memoli2020influenza,hall1991rsv}. Susceptibility remains at its late value afterwards.\\
Baseline vaccine efficacy, \(VE_{S,0}\) & 0.20, 0.50, 0.80 & 0.20, 0.50, 0.80 & 0.20, 0.50, 0.80 & Matched low, intermediate and high efficacy levels selected from Table~\ref{tab:simulation_design}; evidence on variable protection motivates the range~\cite{feikin2022duration}.\\
Transmission coefficient, \(\beta\) & Recalibrated & Recalibrated & Recalibrated & Spectral calibration~\cite{wang2003epidemic,van2008virus} using Eq.~\ref{eq:beta_calibration}, with each profile's \(R_0\) and \(\gamma\).\\
\end{longtable}

\begin{table}[!ht]
\centering\small
\caption{\textbf{IFR values used to calculate expected deaths.} Entries are probabilities, not percentages. COVID-like values follow age-specific IFR estimates~\cite{verity2020estimates}. Influenza-like values combine age-specific symptomatic case-fatality estimates with a symptomatic fraction~\cite{mcdonald2023inference,carrat2008influenza}. RSV-like values are sensitivity proxies informed by adult mortality patterns and older-adult disease-burden estimates~\cite{hansen2022mortality,savic2023burden}. Their construction is described in Section~\ref{sec:disease_robustness}.}
\label{tab:ifr_profiles}
\begin{tabular}{rrrr}
\toprule Representative age & COVID-like & Influenza-like & RSV-like\\ \midrule
15 & 0.0000695 & 0.00000669 & 0.000001\\
24 & 0.000309 & 0.00001338 & 0.000003\\
34.5 & 0.000844 & 0.00002342 & 0.000010\\
44.5 & 0.001610 & 0.00004683 & 0.000030\\
54.5 & 0.005950 & 0.00012042 & 0.000150\\
64.5 & 0.019300 & 0.00049841 & 0.006600\\
75 & 0.042800 & 0.00327810 & 0.012000\\
\bottomrule
\end{tabular}
\end{table}

\subsection{Sensitivity to network structure}
\label{sec:network_robustness}

Network sensitivity uses seven network types (Table~\ref{tab:network_variants}). All retain the same residents, individual attributes and external exposure. Internal weighted network strength, total exposure strength and the high-exposure ranking are recalculated for each network. The transmission coefficient is also recalibrated using that network's weighted spectral radius and internal exposure fraction. Simulations use \(R_0=1,2,4,8\), \(VE_{S,0}=0.20,0.50,0.80\), the COVID-like baseline, 50\% coverage and 100 replicates. Epidemic streams are independent across strategies in this workflow. Random recipient sets are generated separately from epidemic randomness and remain fixed across the replicates of a setting.

\begin{figure}[p]
    \centering
    \includegraphics[width=\linewidth,height=0.7\textheight,keepaspectratio]{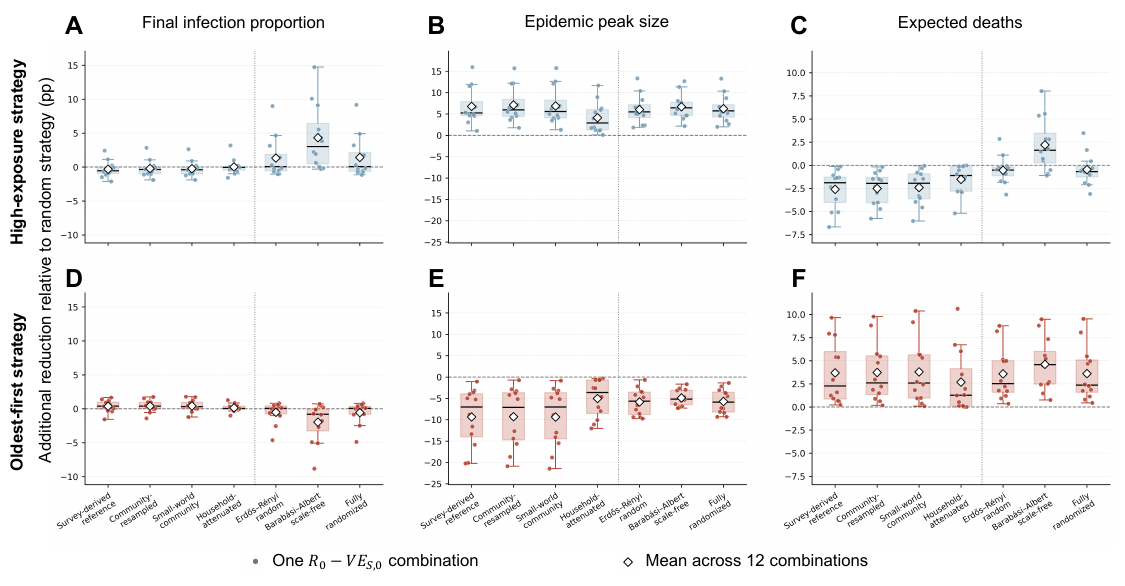}
    \caption{\textbf{Sensitivity of strategy performance to network structure under leaky protection.} Additional reductions relative to the random strategy are shown for the high-exposure (\textit{A--C}) and oldest-first (\textit{D--F}) strategies across seven network types (Table~\ref{tab:network_variants}). Columns show final infection proportion (\textit{A}, \textit{D}), epidemic peak size (\textit{B}, \textit{E}) and expected deaths (\textit{C}, \textit{F}). Each point represents one of 12 combinations of \(R_0=1,2,4,8\) and \(VE_{S,0}=0.20,0.50,0.80\). Values use one reference network, 10 realizations per local perturbation and five realizations per idealized topology. Additional reductions are calculated within each realization and then averaged for each parameter combination. All settings use 50\% coverage and 100 simulations of 500 days per strategy and realization. Box plots summarize the 12 parameter-level values. Centre lines show medians, boxes show interquartile ranges, whiskers extend to the most extreme observations within 1.5 interquartile ranges of the box edges, and diamonds show means. Values are in pp, with positive values favouring the targeted strategy. Horizontal dashed lines mark zero. Vertical dotted lines separate the reference network and its local perturbations from the idealized topologies.}
    \label{fig:network_robustness}
\end{figure}

The reference network has one realization. Each local structural perturbation has 10 realizations, and each idealized topology has five. Additional reductions are calculated using the matching no-vaccination baseline for each realization. They are then averaged over network realizations within each parameter combination. Each distribution in Fig.~\ref{fig:network_robustness} therefore contains 12 parameter-level values, not one observation per network realization.

\begin{longtable}{L{0.20\textwidth}L{0.08\textwidth}L{0.36\textwidth}L{0.27\textwidth}}
\caption{\textbf{Network variants used in the structural sensitivity analysis.} Idealized topologies are controls for structural assumptions, not reconstructions of observed contact networks.}\label{tab:network_variants}\\
\toprule Network & Realizations & Construction & Weight constraint\\ \midrule
\endfirsthead
\multicolumn{4}{l}{\textit{Table~\ref{tab:network_variants} continued.}}\\
\toprule Network & Realizations & Construction & Weight constraint\\ \midrule \endhead
\midrule \multicolumn{4}{r}{\textit{Continued on next page.}}\\ \endfoot
\bottomrule \endlastfoot
Survey-derived reference & 1 & Original household and stochastic community layers & Original weights\\
Community-resampled & 10 & Original households, newly sampled community partners & Same individual activity budgets and encounter weight\\
Small-world community & 10 & Original households, Watts--Strogatz community layer with rewiring probability 0.10 and randomly relabelled nodes & Community edge count matches the reference non-household count, weights sampled from reference and rescaled\\
Household-attenuated & 10 & Household weights halved, community contacts regenerated & Community weights rescaled to recover reference total internal strength\\
Erd\H{o}s--R\'enyi random & 5 & Uniformly sampled distinct edges, reference total edge count & Weights sampled with replacement and rescaled\\
Barab\'asi--Albert scale-free & 5 & Preferential attachment, random node relabelling, edge count adjusted to the reference & Weights sampled with replacement and rescaled\\
Fully randomized & 5 & Uniformly sampled distinct edges, reference total edge count & Reference weights permuted over edges\\
\end{longtable}

For the small-world construction, twice the target community edge count divided by \(N\) is rounded to an integer. If this integer is odd, it is increased by one to obtain an even initial ring degree. Household edges are excluded from the new community edge list. Excess edges are removed and missing edges are sampled to reach the target. For the Barab\'asi--Albert construction, the attachment count is the reference total edge count divided by \(N\), rounded to the nearest integer and bounded below by one. For the idealized topologies, original household edges are not retained. Rescaling matches the total internal strength where specified, but does not preserve each individual's internal strength.

\subsection{Monte Carlo uncertainty}
\label{sec:uncertainty}

For each mean outcome \(Y_m^s\), the Monte Carlo standard error \(SE(Y_m^s)\) is the standard deviation across replicate outcomes divided by \(\sqrt{n}\), where \(n=100\). We report normal-approximation intervals \(Y_m^s\pm1.96SE(Y_m^s)\)~\cite{morris2019simulation}. They describe numerical uncertainty in the mean under the specified model, network and recipient set, not uncertainty in survey sampling, disease parameters or model structure.

\begin{figure}[p]
    \centering
    \includegraphics[width=\linewidth,height=0.7\textheight,keepaspectratio]{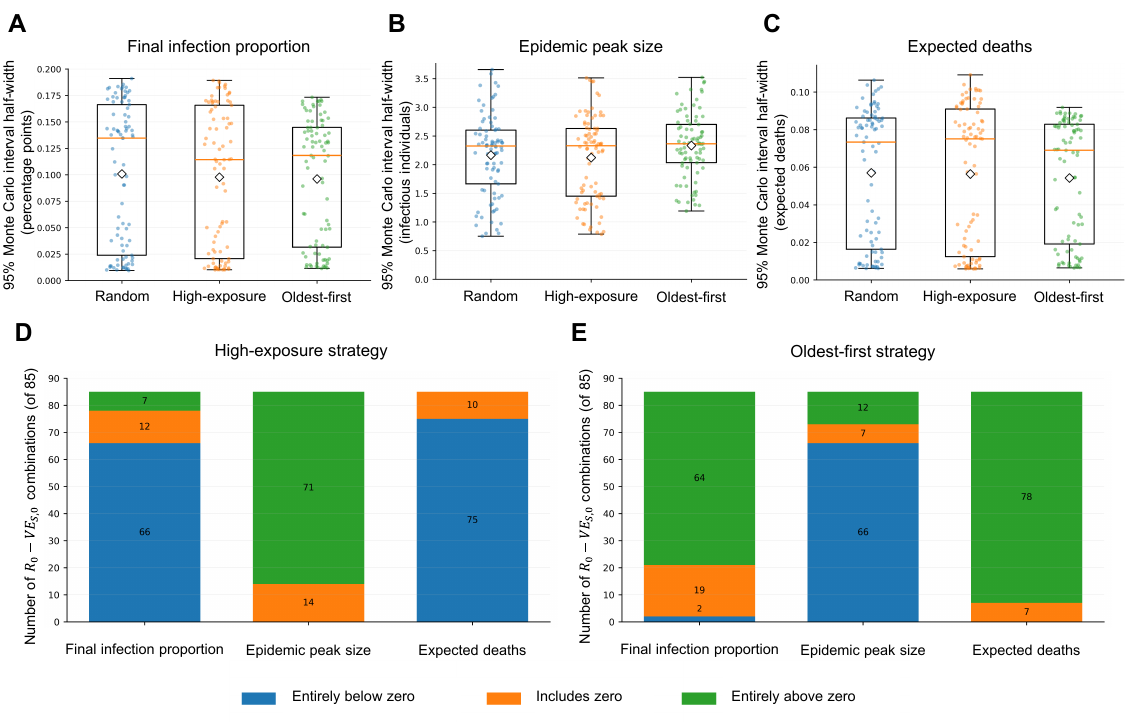}
    \caption{\textbf{Monte Carlo uncertainty across the principal leaky parameter grid.} Panels \textit{A--C} show half-widths of approximate 95\% Monte Carlo intervals for mean final infection proportion (\textit{A}, pp), epidemic peak size (\textit{B}, individuals) and expected deaths (\textit{C}). Results are shown separately for the random, high-exposure and oldest-first strategies. Within each strategy, each point represents one of 85 combinations of \(R_0\) and \(VE_{S,0}\). Box plots show medians and interquartile ranges, with whiskers extending to the most extreme observations within 1.5 interquartile ranges of the box edges. Diamonds show means. Panels \textit{D} and \textit{E} count settings in which intervals for the additional reduction under the high-exposure and oldest-first strategies, respectively, lie below zero, include zero or lie above zero. Positive values favour the targeted strategy. All estimates use 100 simulations of 500 days at 50\% coverage on the reference network, with recipient sets fixed within each setting. Intervals for strategy differences use independent-sample standard errors and treat the no-vaccination mean as a fixed normalization baseline.}
    \label{fig:monte_carlo_summary}
\end{figure}

In the principal leaky heatmap experiments, strategies do not use synchronized epidemic streams. We therefore propagate standard errors as for independent means. Using the same notation as the main text, the standard error of additional reduction is
\[
SE\{\Delta_m(s)\}=
\frac{100}{Y_m^0}
\sqrt{SE(Y_m^{\mathrm{Random}})^2+SE(Y_m^s)^2}.
\]
The no-vaccination mean is treated as a fixed denominator. Its sampling uncertainty is not propagated. Fig.~\ref{fig:monte_carlo_summary} uses these conditional intervals to describe numerical stability, not to perform a multiple-testing procedure across the parameter grid.

For hybrid allocation, each outcome at \(\alpha\) is paired with the same replicate at \(\alpha=0\). We calculate the standard error from the 100 within-replicate differences. The same procedure applies to endpoint decile contributions. Intervals along \(\alpha\) are pointwise, not simultaneous confidence bands. An interval that includes zero does not establish equivalence.

\section{Supplementary Results}

\subsection{Comparison with all-or-nothing protection}
\label{sec:aon_results}

\begin{figure}[!htbp]
    \centering
    \includegraphics[width=\linewidth,height=0.7\textheight,keepaspectratio]{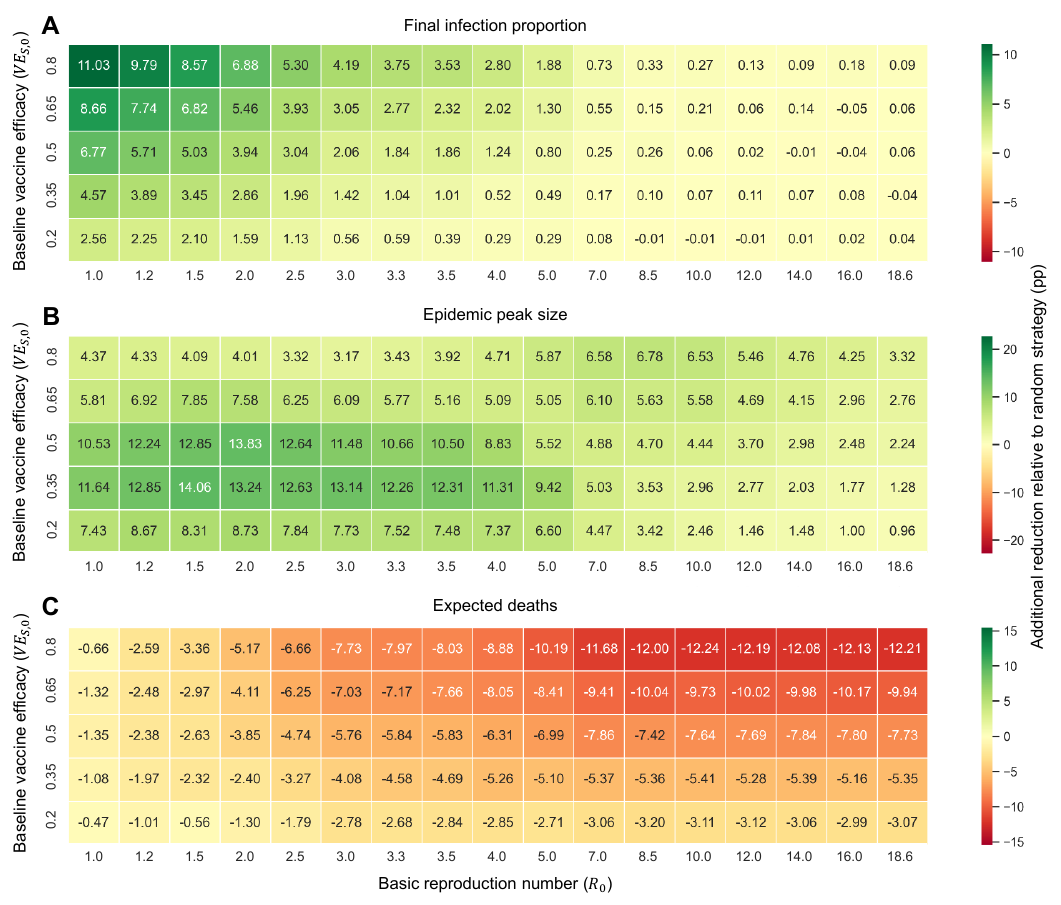}
    \caption{\textbf{Under all-or-nothing protection, the high-exposure strategy consistently reduces epidemic peak size relative to the random strategy.} Heatmaps show additional reductions in final infection proportion (\textit{A}), epidemic peak size (\textit{B}) and expected deaths (\textit{C}) across combinations of \(R_0\) and \(VE_{S,0}\). Positive values favour the high-exposure strategy, whereas negative values favour the random strategy. Additional reductions are expressed in percentage points (pp) using the corresponding no-vaccination mean as the normalization baseline. Results are averaged across 100 simulations of 500 days per setting at 50\% vaccination coverage. All-or-nothing protection follows the waning schedule described in Section~\ref{sec:aon}.}
    \label{fig:aon_high_exposure}
\end{figure}

\begin{figure}[p]
    \centering
    \includegraphics[width=\linewidth,height=0.7\textheight,keepaspectratio]{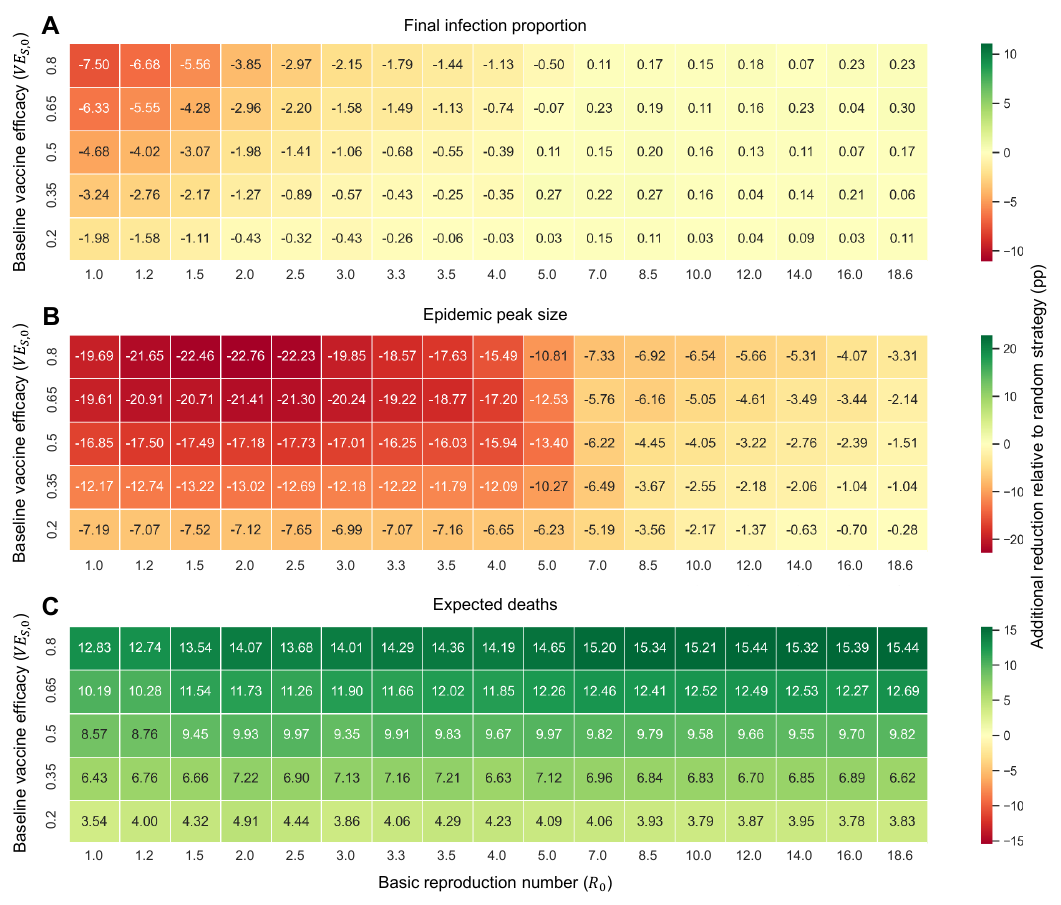}
    \caption{\textbf{Under all-or-nothing protection, the oldest-first strategy reduces expected deaths but produces a larger epidemic peak size than the random strategy.} Heatmaps show additional reductions in final infection proportion (\textit{A}), epidemic peak size (\textit{B}) and expected deaths (\textit{C}). Positive values favour the oldest-first strategy, whereas negative values favour the random strategy. Values are in pp. Results are averaged across 100 simulations of 500 days per setting at 50\% vaccination coverage. Parameter combinations, waning and normalization are the same as in Fig.~\ref{fig:aon_high_exposure}.}
    \label{fig:aon_oldest_first}
\end{figure}

\begin{figure}[p]
    \centering
    \includegraphics[width=\linewidth,height=0.7\textheight,keepaspectratio]{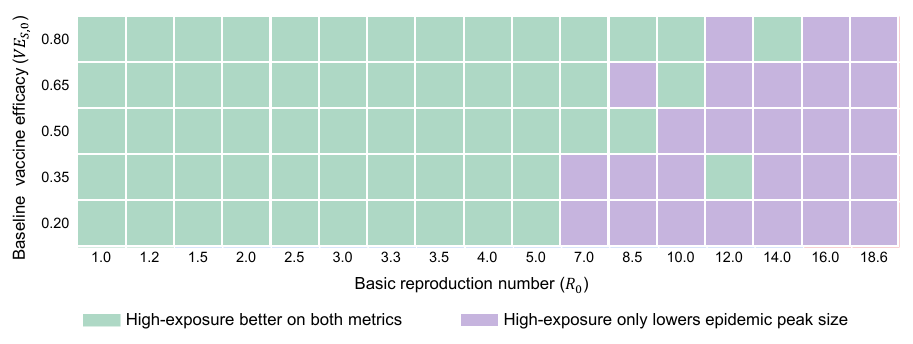}
    \caption{\textbf{Endpoint rankings of the two targeted strategies under all-or-nothing protection.} Colours compare mean final infection proportion and epidemic peak size at 50\% vaccination coverage. Green indicates lower values of both metrics under the high-exposure strategy. Purple indicates a lower epidemic peak size under the high-exposure strategy but a slightly lower final infection proportion under the oldest-first strategy. Classification uses unrounded means from the 100 simulations per setting in Figs.~\ref{fig:aon_high_exposure} and \ref{fig:aon_oldest_first}. It describes endpoint rankings rather than the response along hybrid allocation, which is examined in Fig.~\ref{fig:aon_leaky_hybrid}.}
    \label{fig:aon_endpoint}
\end{figure}

With the same waning schedule, all-or-nothing and leaky protection produce different patterns of strategy performance. In the all-or-nothing full-grid comparison, the high-exposure strategy reduces epidemic peak size more than the random strategy in all 85 settings. Its advantage in final infection proportion is largest at low transmission. At higher transmission, the difference from the random strategy becomes small and is negative in some settings (Fig.~\ref{fig:aon_high_exposure}). The oldest-first strategy reduces expected deaths more than the random strategy throughout the grid but provides weaker peak control (Fig.~\ref{fig:aon_oldest_first}). All-or-nothing protection therefore does not make one strategy preferable for all three metrics.

The all-or-nothing endpoint map contains 59 settings with a high-exposure advantage in both infection metrics. The other 26 have a lower peak under the high-exposure strategy but a slightly lower final infection proportion under the oldest-first strategy (Fig.~\ref{fig:aon_endpoint}). In these purple cells, the high-exposure minus oldest-first difference in final infection proportion ranges from 0.018 to 0.233 pp, with a median of 0.088 pp. Peak size is lower by 11.38--83.73 individuals. The map describes endpoint rankings and effect sizes. It does not establish that final infection proportion rises consistently as \(\alpha\) increases.

\begin{figure}[p]
    \centering
    \includegraphics[width=\linewidth,height=0.7\textheight,keepaspectratio]{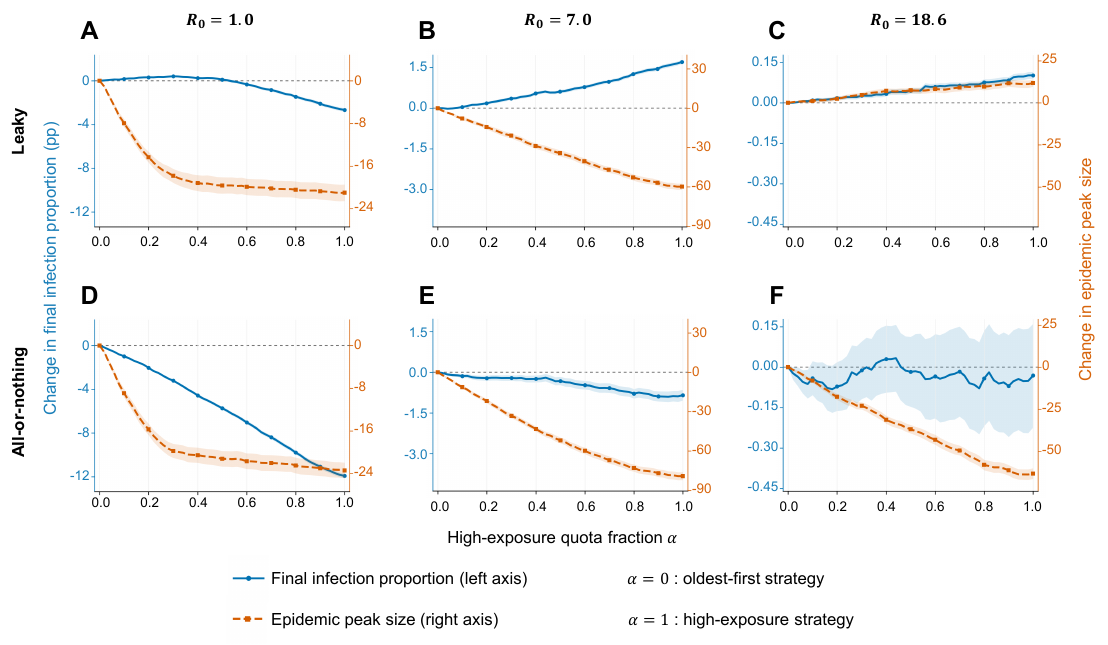}
    \caption{\textbf{Responses to increasing exposure prioritization under leaky and all-or-nothing protection.} Rows show leaky (\textit{A--C}) and all-or-nothing (\textit{D--F}) protection. Columns correspond to \(R_0=1.0\), 7.0 and 18.6. All scenarios use \(VE_{S,0}=0.80\) and 50\% vaccination coverage. Common recipients remain vaccinated, while \(\alpha\) specifies the share of the remaining vaccination places assigned to the high-exposure strategy. Blue lines show changes in final infection proportion (pp, left axes), and orange dashed lines show changes in epidemic peak size (individuals, right axes). Changes are relative to the oldest-first endpoint within each model. Points show means across 100 simulations at each of 51 equally spaced values of \(\alpha\). Shading shows pointwise 95\% Monte Carlo intervals calculated from paired differences. Scales for each metric match across rows within a column. Line crossings on the two vertical axes have no quantitative meaning.}
    \label{fig:aon_leaky_hybrid}
\end{figure}

The matched hybrid simulations help distinguish these patterns (Fig.~\ref{fig:aon_leaky_hybrid} and Table~\ref{tab:aon_leaky_endpoints}). At \(R_0=1.0\), both models favour the high-exposure endpoint for both infection metrics. At \(R_0=7.0\), increasing exposure prioritization under leaky protection produces a clear overall increase in final infection proportion and a decrease in epidemic peak size. Under all-or-nothing protection, both endpoint differences remain negative. At \(R_0=18.6\), the leaky high-exposure endpoint has slightly higher values for both metrics. Under all-or-nothing protection, epidemic peak size decreases substantially, whereas final infection proportion changes little and has no clear sustained increase over \(\alpha\).

For the high-transmission all-or-nothing scenario, the range of the 51 mean infection proportions is 0.115 pp. The endpoint difference is \(-0.031\) pp, with a 95\% Monte Carlo interval from \(-0.224\) to 0.161 pp. Its peak difference is \(-63.20\) individuals, with an interval from \(-66.39\) to \(-60.01\). This supports a pronounced response in epidemic peak size with little change in final infection proportion in this scenario. 

The full-grid and hybrid results come from separate simulation batches. A small endpoint difference can therefore change sign between them, as occurs for all-or-nothing at \(R_0=18.6\) and \(VE_{S,0}=0.80\). The full-grid map uses the data in Figs.~\ref{fig:aon_high_exposure} and \ref{fig:aon_oldest_first}, while the hybrid curves show their own paired estimates.

These comparisons show a clearer change in the relationship between the two infection metrics under leaky protection in the representative scenarios. The endpoint regions also have descriptive boundaries rather than statistically estimated transition points. 

\begin{table}[!ht]
\centering\small
\caption{\textbf{Paired endpoint differences in the matched hybrid simulations.} Entries compare the high-exposure endpoint with the oldest-first endpoint within each model. Brackets contain pointwise 95\% Monte Carlo intervals from 100 paired replicates. \(VE_{S,0}=0.80\) and vaccination coverage is 50\%. Negative differences favour the high-exposure endpoint.}
\label{tab:aon_leaky_endpoints}
\begin{tabular}{l r L{0.33\textwidth}L{0.33\textwidth}}
\toprule Model & \(R_0\) & Final infection proportion, pp & Epidemic peak size, individuals\\ \midrule
Leaky & 1.0 & \(-2.675\) [\(-2.845,-2.506\)] & \(-21.01\) [\(-22.56,-19.46\)]\\
Leaky & 7.0 & 1.712 [1.637, 1.786] & \(-59.77\) [\(-62.22,-57.32\)]\\
Leaky & 18.6 & 0.102 [0.090, 0.115] & 11.77 [9.52, 14.02]\\
All-or-nothing & 1.0 & \(-11.925\) [\(-12.138,-11.711\)] & \(-23.43\) [\(-24.89,-21.97\)]\\
All-or-nothing & 7.0 & \(-0.846\) [\(-1.048,-0.645\)] & \(-79.09\) [\(-82.12,-76.06\)]\\
All-or-nothing & 18.6 & \(-0.031\) [\(-0.224,0.161\)] & \(-63.20\) [\(-66.39,-60.01\)]\\
\bottomrule
\end{tabular}
\end{table}

\subsection{Sensitivity to vaccination coverage}

\begin{figure}[p]
    \centering
    \includegraphics[width=\linewidth,height=0.7\textheight,keepaspectratio]{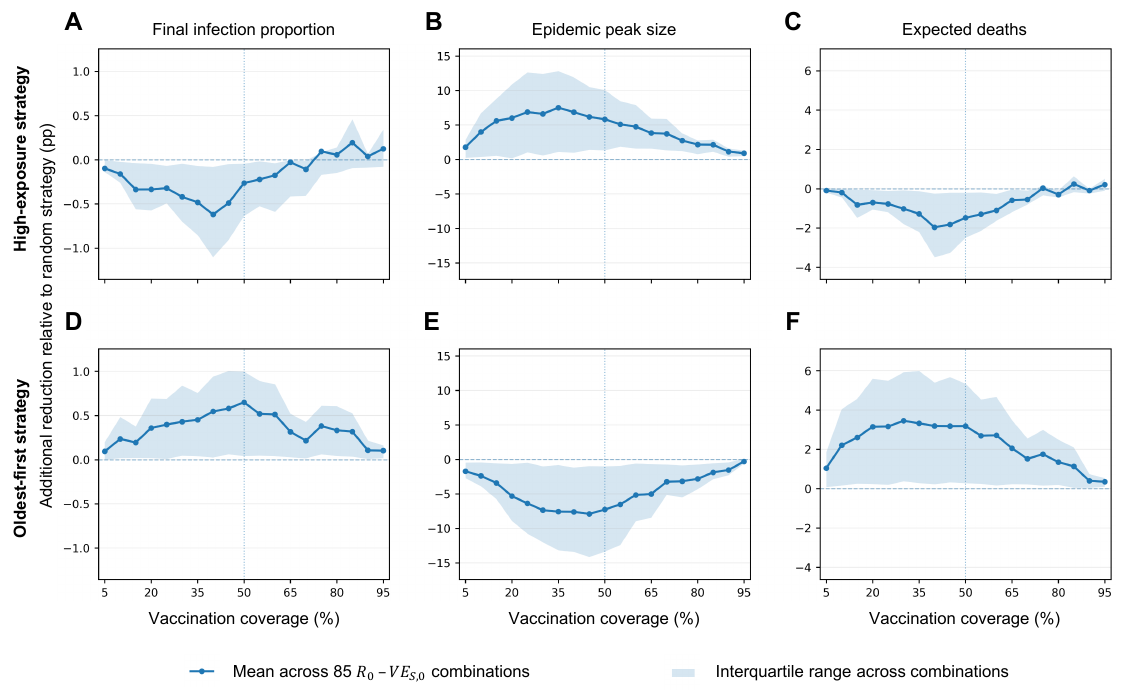}
    \caption{\textbf{Strategy performance across vaccination coverage levels under leaky protection.} Additional reductions relative to the random strategy are shown for the high-exposure (\textit{A--C}) and oldest-first (\textit{D--F}) strategies. Columns show final infection proportion (\textit{A}, \textit{D}), epidemic peak size (\textit{B}, \textit{E}) and expected deaths (\textit{C}, \textit{F}). Points and lines show the mean across 85 combinations of \(R_0\) and \(VE_{S,0}\) at each coverage from 5\% to 95\% in 5 pp increments. Shading shows the interquartile range across parameter combinations, not Monte Carlo uncertainty. Each combination uses 100 simulations of 500 days per strategy. Additional reductions are in pp and use the matching no-vaccination mean as the normalization baseline. Positive values favour the targeted strategy. The horizontal dashed line marks zero, and the vertical dotted line marks 50\% coverage.}
    \label{fig:coverage_robustness}
\end{figure}

Across the coverage sweep, the clearest average benefit of the high-exposure strategy remains peak reduction, whereas that of the oldest-first strategy remains reduction in expected deaths (Fig.~\ref{fig:coverage_robustness}). At 50\% coverage, the high-exposure mean additional reductions across the 85 settings are \(-0.26\) pp for final infection proportion, 5.83 pp for epidemic peak size and \(-1.48\) pp for expected deaths. The corresponding oldest-first values are 0.65, \(-7.27\) and 3.18 pp. These are equally weighted averages across the selected parameter grid, not population-level estimates for a specific disease.

\subsection{Sensitivity to disease parameters}

\begin{figure}[p]
    \centering
    \includegraphics[width=\linewidth,height=0.7\textheight,keepaspectratio]{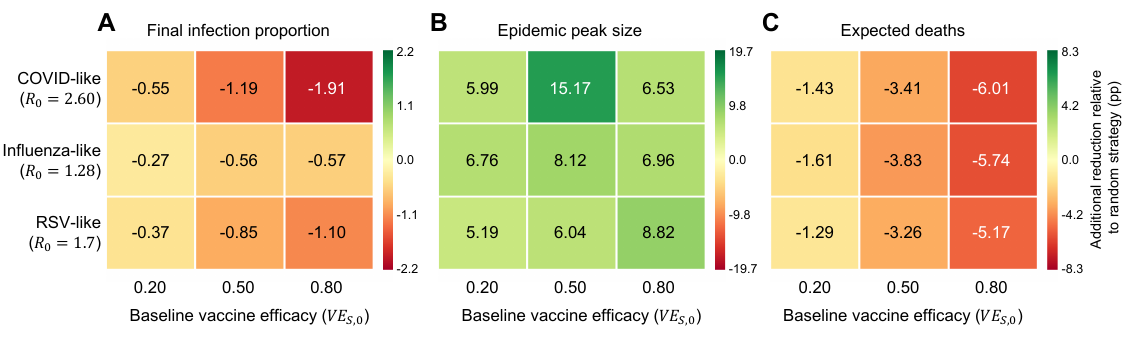}
    \caption{\textbf{Performance of the high-exposure strategy under alternative disease parameterizations.} Heatmaps show additional reductions relative to the random strategy under leaky protection for final infection proportion (\textit{A}), epidemic peak size (\textit{B}) and expected deaths (\textit{C}). Rows correspond to COVID-like, influenza-like and RSV-like scenarios, with reference \(R_0\) values of 2.60, 1.28 and 1.70, respectively. Columns correspond to \(VE_{S,0}=0.20\), 0.50 and 0.80. Disease-course and immunity parameters are listed in Table~\ref{tab:disease_parameters}, and each scenario uses its own IFR profile (Table~\ref{tab:ifr_profiles}). The internal contact network and individual external exposure components remain fixed. Positive values favour the high-exposure strategy, whereas negative values favour the random strategy. Values are in pp and use the corresponding no-vaccination mean as the normalization baseline. Results are averaged across 100 simulations of 500 days at 50\% coverage.}
    \label{fig:disease_high_exposure}
\end{figure}

\begin{figure}[p]
    \centering
    \includegraphics[width=\linewidth,height=0.7\textheight,keepaspectratio]{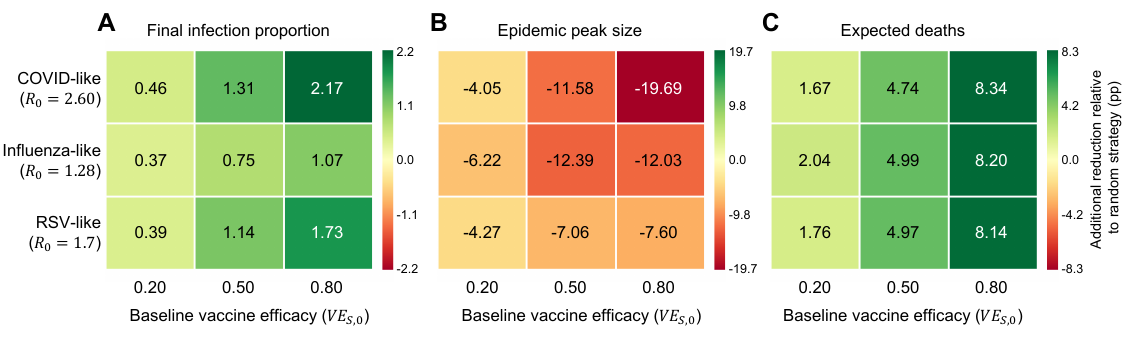}
    \caption{\textbf{Performance of the oldest-first strategy under alternative disease parameterizations.} Heatmaps show additional reductions relative to the random strategy under leaky protection for final infection proportion (\textit{A}), epidemic peak size (\textit{B}) and expected deaths (\textit{C}). Rows correspond to COVID-like, influenza-like and RSV-like scenarios, with \(R_0=2.60\), 1.28 and 1.70, respectively. Columns correspond to \(VE_{S,0}=0.20\), 0.50 and 0.80. Each scenario uses the disease-course, immunity and IFR parameters in Tables~\ref{tab:disease_parameters} and \ref{tab:ifr_profiles}. Positive values favour the oldest-first strategy, whereas negative values favour the random strategy. Values are in pp. Results are averaged across 100 simulations of 500 days at 50\% coverage. Network settings and normalization are the same as in Fig.~\ref{fig:disease_high_exposure}.}
    \label{fig:disease_oldest_first}
\end{figure}

The alternative disease profiles retain the contrast between peak control and mortality reduction (Figs.~\ref{fig:disease_high_exposure} and \ref{fig:disease_oldest_first}). In all nine settings, the high-exposure strategy improves peak reduction relative to the random strategy but performs worse for final infection proportion and expected deaths. The oldest-first strategy has the opposite signs. These tests support objective-dependent performance under the selected alternative profiles. Because each disease profile uses one reference \(R_0\), they do not establish all three transmission patterns within every pathogen.

\subsection{Sensitivity to network structure}

Across all seven network types, the high-exposure strategy provides greater peak reduction than the random strategy in each of the 12 parameter combinations after averaging results across network realizations. The oldest-first strategy provides greater reduction in expected deaths in every such comparison (Fig.~\ref{fig:network_robustness}). The reference network and three local perturbations also retain negative additional reductions in expected deaths for the high-exposure strategy.

Final infection proportion is more sensitive to topology. In the Barab\'asi--Albert control, the high-exposure mean additional reduction reaches 4.33 pp for final infection proportion and 2.21 pp for expected deaths. The mean additional reduction in final infection proportion is negative for the oldest-first strategy. These changes show that preserving broad peak and mortality advantages does not imply invariance of every strategy ranking. The idealized controls alter how internal contact opportunities are distributed across residents and should not be interpreted as equivalent empirical populations.

\subsection{Numerical uncertainty in the principal comparisons}

The Monte Carlo intervals support the main performance differences while identifying settings with weak preferences (Fig.~\ref{fig:monte_carlo_summary}). For the high-exposure strategy, intervals for the additional reduction in epidemic peak size lie above zero in 71 of 85 settings and include zero in the other 14. The corresponding intervals for final infection proportion lie below zero in 66 settings, include zero in 12 and lie above zero in seven. Those for expected deaths lie below zero in 75 settings and include zero in 10.

For the oldest-first strategy, intervals for the additional reduction in expected deaths lie above zero in 78 settings and include zero in seven. The corresponding intervals for epidemic peak size lie below zero in 66 settings, include zero in seven and lie above zero in 12. Those for final infection proportion lie above zero in 64 settings, include zero in 19 and lie below zero in two. These are descriptive counts of conditional Monte Carlo intervals, not multiplicity-adjusted discoveries.

\begin{figure}[p]
    \centering
    \includegraphics[width=\linewidth,height=0.7\textheight,keepaspectratio]{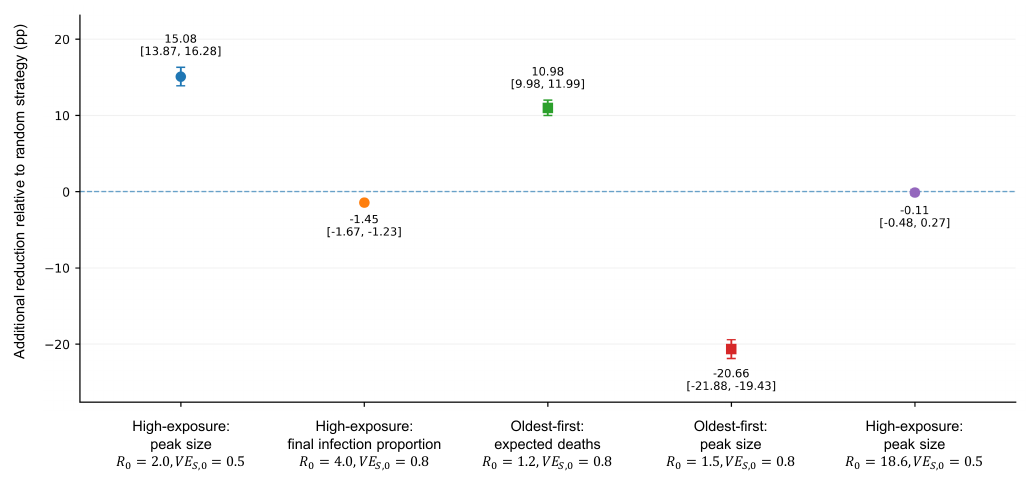}
    \caption{\textbf{Monte Carlo uncertainty for representative strategy comparisons under leaky protection.} Points and error bars show additional reductions relative to the random strategy and approximate 95\% Monte Carlo intervals for five selected comparisons. Parameters are shown below each estimate. All values are in pp and use the corresponding no-vaccination mean as the normalization baseline. Each estimate uses 100 simulations of 500 days per strategy at 50\% coverage. Intervals combine the independent-sample standard errors of the targeted and random strategies, with the no-vaccination mean treated as fixed. The horizontal dashed line marks zero.}
    \label{fig:representative_uncertainty}
\end{figure}

The representative estimates illustrate the difference between large and near-zero effects. At \(R_0=2.0\) and \(VE_{S,0}=0.50\), the high-exposure additional peak reduction is 15.08 pp, with an interval of 13.87--16.28 pp. At \(R_0=18.6\) and \(VE_{S,0}=0.50\), it is \(-0.11\) pp, with an interval of \(-0.48\)--0.27 pp. The latter interval includes zero, although the point estimate favours the random strategy (Fig.~\ref{fig:representative_uncertainty}).

\clearpage
\section*{Supplementary Figures}

\clearpage

\clearpage

\FloatBarrier
\bibliographystyle{unsrtnat}
\bibliography{leaky_vaccine_SI}